%% file: ms.tex
\documentclass[
a4paper,     
12pt,        
article,
onecolumn,   
openany,     
]{memoir}

\usepackage[utf8]{inputenc}

\usepackage[T1]{fontenc}
\usepackage{lmodern}

\usepackage[english]{babel}

\usepackage[final]{microtype}

\usepackage{amsmath,amssymb,mathtools}

\usepackage{pifont}

\usepackage{graphicx}
\usepackage{makeidx}
\usepackage{import}
\usepackage{algorithm}
\usepackage{algpseudocode}

\algdef{SE}[SUBALG]{Indent}{EndIndent}{}{\algorithmicend\ }

\algtext*{Indent}
\algtext*{EndIndent}

\algnewcommand{\LineComment}[1]{\State \(\triangleright\) #1}

\usepackage[top=1in, bottom=1.25in, inner=1in, outer=1in]{geometry}

\newlength\forceindent
\maxsecnumdepth{paragraph}
\setsecnumdepth{paragraph}

\usepackage{titlesec}
\usepackage{etoolbox}

\titlespacing\section{0pt}{12pt plus 4pt minus 2pt}{0pt plus 2pt minus 2pt}
\titlespacing\subsection{0pt}{12pt plus 2pt minus 1pt}{0pt plus 1pt minus 1pt}
\titlespacing\subsubsection{0pt}{12pt plus 2pt minus 1pt}{0pt plus 1pt minus 1pt}

\counterwithout{section}{chapter}
\titleformat*{\section}{\Large\bfseries}

\setsecheadstyle{\normalfont\large\bfseries}
\setsubsecheadstyle{\normalfont\normalsize\bfseries}
\setparaheadstyle{\normalfont\normalsize\bfseries}
\setparaindent{0pt}\setafterparaskip{0pt}

\captiondelim{\space}
\captionnamefont{\small\bfseries}
\captiontitlefont{\small\normalfont}

\changecaptionwidth        
\captionwidth{1\textwidth} 

\usepackage{tabularx}

\makepagestyle{memoirStylePages}
\makerunningwidth{memoirStylePages}{\textwidth}

\makeheadrule{memoirStylePages}{\textwidth}{\normalrulethickness}
\makefootrule{memoirStylePages}{\textwidth}{\normalrulethickness}{0pt}

\makeevenfoot{memoirStylePages}{}{\bfseries\thepage}{}
\makeoddfoot{memoirStylePages}{}{\bfseries\thepage}{}
\makeevenhead{memoirStylePages}{}{\textit{ESCAPE - WP1: Weather \& Climate Dwarfs - D1.2: Batch no. 2 of Dwarfs}}{}
\makeoddhead{memoirStylePages}{}{\textit{ESCAPE - WP1: Weather \& Climate Dwarfs - D1.2: Batch no. 2 of Dwarfs}}{}

\makeatletter
\makepsmarks{memoirStylePages}{
\createmark{chapter}{both}{shownumber}{\@chapapp\ }{ \quad }
\createmark{section}{right}{shownumber}{}{ \quad }
\createplainmark{toc}{both}{\contentsname}
\createplainmark{lof}{both}{\listfigurename}
\createplainmark{lot}{both}{\listtablename}
\createplainmark{bib}{both}{\bibname}
\createplainmark{index}{both}{\indexname}
\createplainmark{glossary}{both}{\glossaryname}
}
\makeatother

\aliaspagestyle{chapter}{chap}

\usepackage{pdfpages}

\usepackage{biblatex}
\usepackage{xspace}
\usepackage{xcolor}
\usepackage{tikz}
\definecolor{gray}{rgb}{0.4,0.4,0.4}
\definecolor{lightgray}{rgb}{0.9,0.9,0.9}
\definecolor{darkblue}{rgb}{0.0,0.0,0.6}
\definecolor{cyan}{rgb}{0.0,0.6,0.6}
\definecolor{maroon}{rgb}{0.5,0.0,0.0}
\definecolor{darkgreen}{rgb}{0.0,0.5,0.0}

\usepackage{listings}
\usepackage{lstautogobble}

\lstdefinestyle{BashStyle}
{
  language=bash,
  basicstyle=\footnotesize\ttfamily,
  frame=single,
  columns=fullflexible,
  backgroundcolor=\color{yellow!10},
  linewidth=\linewidth,
  xleftmargin=0.05\linewidth,
  keepspaces=true,
  framesep=5pt,
  rulecolor=\color{black!30},
  aboveskip=10pt,
  autogobble=true
}

\lstdefinelanguage{XML}
{
  basicstyle=\ttfamily\footnotesize,
  morestring=[b]",
  moredelim=[s][\bfseries\color{maroon}]{<}{\ },
  moredelim=[s][\bfseries\color{maroon}]{</}{>},
  moredelim=[l][\bfseries\color{maroon}]{/>},
  moredelim=[l][\bfseries\color{maroon}]{>},
  morecomment=[s]{<?}{?>},
  morecomment=[s]{<!--}{-->},
  commentstyle=\color{gray},
  stringstyle=\color{orange},
  identifierstyle=\color{darkblue},
  showstringspaces=false
}

\lstdefinestyle{XMLStyle}
{
  language=make,
  basicstyle=\ttfamily\footnotesize,
  numbers=left,
  numberstyle=\tiny,
  numbersep=3pt,
  frame=,
  columns=fullflexible,
  backgroundcolor=\color{black!05},
  linewidth=\linewidth,
  xleftmargin=0.05\linewidth,
  keepspaces=true
}
\lstdefinestyle{CStyle}{
  belowcaptionskip=1\baselineskip,
  breaklines=true,
  frame=single,
  escapeinside={\%*}{*)},
  xleftmargin=\parindent,
  language=C,
  captionpos=b,
  keepspaces=true,
  backgroundcolor=\color{black!05},
  showstringspaces=false,
  numbers=left,
  numbersep=5pt,
  numberstyle=\tiny\color{black},
  basicstyle=\scriptsize\ttfamily,
  keywordstyle=\bfseries\color{green!40!black},
  commentstyle=\itshape\color{purple!40!black},
  identifierstyle=\color{blue},
  stringstyle=\color{orange},
  tabsize=4
}

\lstdefinestyle{CStyleNoLine}{
  belowcaptionskip=1\baselineskip,
  breaklines=true,
  frame=single,
  escapeinside={\%*}{*)},
  xleftmargin=\parindent,
  language=C,
  captionpos=b,
  keepspaces=true,
  backgroundcolor=\color{black!05},
  showstringspaces=false,
  basicstyle=\scriptsize\ttfamily,
  keywordstyle=\bfseries\color{green!40!black},
  commentstyle=\itshape\color{purple!40!black},
  identifierstyle=\color{blue},
  stringstyle=\color{orange},
  tabsize=4
}

\lstdefinestyle{FStyle}{
  belowcaptionskip=1\baselineskip,
  breaklines=true,
  frame=single,
  escapeinside={\%*}{*)},
  xleftmargin=\parindent,
  language=[90]Fortran,
  captionpos=b,
  keepspaces=true,
  backgroundcolor=\color{red!05},
  showstringspaces=false,
  numbers=left,
  numbersep=5pt,
  numberstyle=\tiny\color{black},
  basicstyle=\scriptsize\ttfamily,
  keywordstyle=\bfseries\color{red!40!black},
  commentstyle=\itshape\color{green!40!black},
  identifierstyle=\color{blue},
  stringstyle=\color{orange},
  tabsize=4
}

\lstdefinestyle{FStyleNoLine}{
  belowcaptionskip=1\baselineskip,
  breaklines=true,
  frame=single,
  escapeinside={\%*}{*)},
  xleftmargin=\parindent,
  language=[90]Fortran,
  captionpos=b,
  keepspaces=true,
  backgroundcolor=\color{red!05},
  showstringspaces=false,
  basicstyle=\scriptsize\ttfamily,
  keywordstyle=\bfseries\color{red!40!black},
  commentstyle=\itshape\color{green!40!black},
  identifierstyle=\color{blue},
  stringstyle=\color{orange},
  tabsize=4
}

\ifdefined\HCode

\else

\fi

\ifdefined\HCode
\newcommand{\inlsh}[1]{\texttt{#1}}
\else
\newcommand{\inlsh}[1]{\tikz[anchor=base,baseline]\node[inner sep=2pt,
outer sep=0,draw=yellow!10,fill=yellow!10]{\texttt{#1}};}
\fi

\usepackage{environ}
\usepackage[tikz]{bclogo}
\usetikzlibrary{calc}

\ifdefined\HCode
\NewEnviron{notebox}{\textbf{Note:} \BODY}
\NewEnviron{warningbox}{\textbf{Warning:} \BODY}
\NewEnviron{tipbox}{\textbf{Tip:} \BODY}
\NewEnviron{custombox}[3]{\textbf{#1} \BODY}
\else
\NewEnviron{notebox}
  {\par\medskip\noindent
  \begin{tikzpicture}
    \node[inner sep=5pt,fill=black!10,draw=black!30] (box)
    {\parbox[t]{.99\linewidth}{%
      \begin{minipage}{.1\linewidth}
      \centering\tikz[scale=1]\node[scale=1.5]{\bcinfo};
      \end{minipage}%
      \begin{minipage}{.85\linewidth}
      \textbf{Note}\par\smallskip
      \BODY
      \end{minipage}\hfill}%
    };
   \end{tikzpicture}\par\medskip%
}
\NewEnviron{warningbox}
  {\par\medskip\noindent
  \begin{tikzpicture}
    \node[inner sep=5pt,fill=red!10,draw=black!30] (box)
    {\parbox[t]{.99\linewidth}{%
      \begin{minipage}{.1\linewidth}
      \centering\tikz[scale=1]\node[scale=1.5]{\bcdanger};
      \end{minipage}%
      \begin{minipage}{.85\linewidth}
      \textbf{Warning}\par\smallskip
      \BODY
      \end{minipage}\hfill}%
    };
   \end{tikzpicture}\par\medskip%
}
\NewEnviron{tipbox}
  {\par\medskip\noindent
  \begin{tikzpicture}
    \node[inner sep=5pt,fill=green!10,draw=black!30] (box)
    {\parbox[t]{.99\linewidth}{%
      \begin{minipage}{.1\linewidth}
      \centering\tikz[scale=1]\node[scale=1.5]{\bclampe};
      \end{minipage}%
      \begin{minipage}{.85\linewidth}
      \textbf{Tip}\par\smallskip
      \BODY
      \end{minipage}\hfill}%
    };
   \end{tikzpicture}\par\medskip%
}
\NewEnviron{custombox}[3]
  {\par\medskip\noindent
  \begin{tikzpicture}
    \node[inner sep=5pt,fill=#3!10,draw=black!30] (box)
    {\parbox[t]{.99\linewidth}{%
      \begin{minipage}{.1\linewidth}
      \centering\tikz[scale=1]\node[scale=1.5]{#2};
      \end{minipage}%
      \begin{minipage}{.85\linewidth}
      \textbf{#1}\par\smallskip
      \BODY
      \end{minipage}\hfill}%
    };
   \end{tikzpicture}\par\medskip%
}
\fi

\makeindex

\usepackage[linktoc=all,hyperfootnotes=false,bookmarksdepth=3]{hyperref}
\hypersetup{
colorlinks,
citecolor=darkblue,
filecolor=darkblue,
linkcolor=darkblue,
urlcolor=darkblue,
pdfborder={0 0 0},
pdfauthor={I am the Author}
}
\usepackage{memhfixc}

\ifdefined\HCode
\else
\makeatletter
\newlength\drop

\newcommand*{\titlepage}{
    \thispagestyle{empty}
    \begingroup
    \drop = 0.1\textheight
    \vspace*{\baselineskip}
    \vfill
    \hbox{
      \hspace*{0.1\textwidth}
      \rule{1pt}{\dimexpr\textheight-28pt\relax}
      \hspace*{0.05\textwidth}
      \parbox[b]{0.85\textwidth}{
        \vbox{
          \vspace{\drop}
          {\LARGE\bfseries\raggedright\@title\par}
          \vskip4\baselineskip
          {\HUGE\bfseries \textcolor{darkblue}{ESCAPE}\par}
          \vskip1.0\baselineskip
          {\large\bfseries\@date\par}
          \vspace{0.1\textheight}
          {\small\noindent\@author}\\[\baselineskip]
        }
      }
    }
    \vfill
    \null
\endgroup}
\makeatother
\fi

\begin{document}

\title{Energy-efficient Scalable Algorithms for Weather Prediction at Exascale}
\author{Research and Innovation Action \newline
H2020-FETHPC-2014 \newline
Author: Daniel Thiemert \newline
Date: \today \newline
Project Coordinator: Dr. Peter Bauer (ECMWF) \newline
Project Start Date: 01/10/2015 \newline
Project Duration: 36 months \newline
Published by the ESCAPE Consortium \newline
Version: 0.1 \newline
Contractual Delivery Date: 30/06/2016 \newline
Work Package/ Task: WP1/ T1.1 \newline
Document Owner: ECMWF \newline
Dissemination level: Public \newline
Contributors: Gianmarco Mengaldo, Willem Deconinck, Michail Diamantakis, Alastair McKinstrey, Piet Termonia, Peter Bauer, Nils Wedi }

\frontmatter

\includepdf[pages={1,2}]{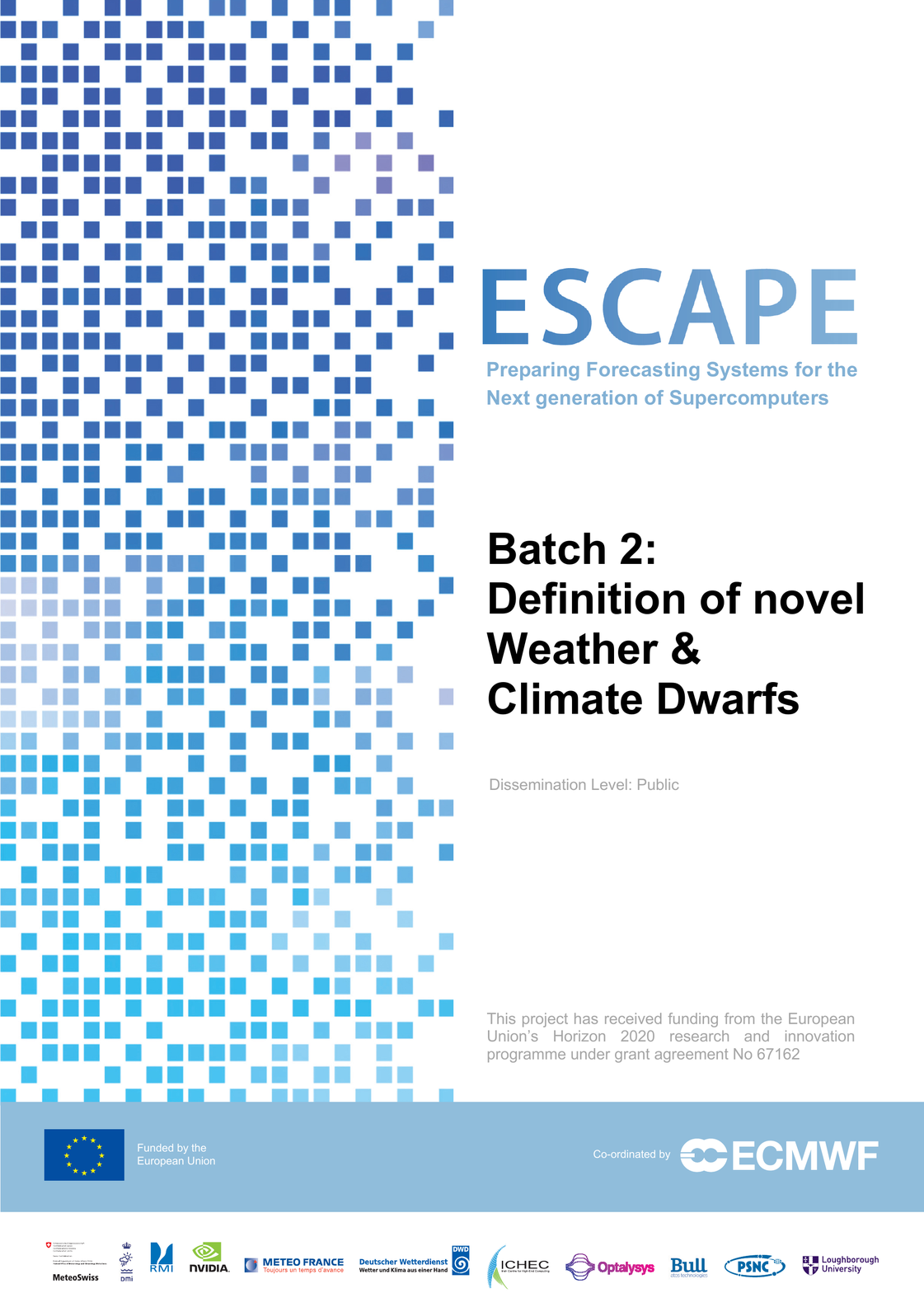}
\setcounter{page}{1}


\setcounter{tocdepth}{2}
\ifx\HCode\undefined
\tableofcontents*
\fi

\mainmatter

\section{\label{ES}Executive summary}
This deliverable contains the description of the characteristics of a second set of so-called numerical weather \& climate prediction dwarfs that form key functional components of prediction models in terms of the science that they encapsulate and in terms of computational cost they impose on the forecast production. The ESCAPE work flow between work packages centres on these dwarfs and hence their selection, their performance assessment, code adaptation and optimisation is crucial for the success of the project. These new dwarfs have been chosen with the purpose of extending the range of computational characteristic represented by the dwarfs previously selected in batch 1 (see Deliverable D1.1). The dwarfs have been made, their documentation has been compiled and the software has been made available on the software exchange platform.

The dwarfs in this deliverable include a multigrid elliptic solver, a novel advection scheme for unstructured meshes, an advection scheme for structured meshes and a radiation scheme. This deliverable includes their scientific description and the guidance for installation, execution and testing. This documentation is equivalent to the one available from the ESCAPE Confluence web-page that disseminates the project outcomes.

\section{\label{IN}Introduction}
\subsection{\label{BG}Background}
ESCAPE stands for Energy-efficient Scalable Algorithms for Weather Prediction at Exascale. The project develops world-class, extreme-scale computing capabilities for European operational numerical weather prediction and future climate models. ESCAPE addresses the ETP4HPC Strategic Research Agenda 'Energy and resiliency' priority topic, promoting a holistic understanding of energy-efficiency for extreme-scale applications using heterogeneous architectures, accelerators and special compute units by:
\begin{itemize}
\item Defining and encapsulating the fundamental algorithmic building blocks underlying weather and climate computing;
\item Combining cutting-edge research on algorithm development for use in extreme-scale, high-performance computing applications, minimising time- and cost-to-solution; 
\item Synthesising the complementary skills of leading weather forecasting consortia, university research, high-performance computing centres, and innovative hardware companies.
\end{itemize}
ESCAPE is funded by the European Commission's Horizon 2020 funding framework under the Future and Emerging Technologies - High-Performance Computing call for research and innovation actions issued in 2014.

\subsection{\label{SC}Scope of this deliverable}
\subsubsection{\label{OB}Objectives of this deliverable}
This document accompanies the prototype implementations of the second batch of weather and climate dwarfs developed in WP1. The dwarfs are used by WPs 2, 3, and 4 for code adaptation, hybrid computing and benchmarking and diagnostics. The document aims to provide a documentation of the provided dwarfs to ensure that they are easily usable by the respective partners.
The dwarf implementations are available at:
\url{https://git.ecmwf.int/projects/ESCAPE}.

\subsubsection{\label{WK}Work performed in this deliverable}
As per the task description in the Description of Action for task 1.1, the work performed in this deliverable included the isolation and packaging of a second set of canonical NWP algorithms and internal model workflows, including:
\begin{itemize}
\item Multigrid elliptic solver (\textit{dwarf-D-ellipticSolver-GCR-multigrid}, see Section \ref{DOC_multigrid})
\item MPDATA for unstructured meshes (\textit{dwarf-D-advection-MPDATA}, see Section \ref{DOC_MPDATA_unstructured})
\item MPDATA for structured meshes (\textit{dwarf-D-advection-MPDATA-structured}, see Section \ref{DOC_MPDATA_structured})
\item ACRANEB2 radiation scheme (\textit{dwarf-P-radiation-ACRANEB2}, see Section \ref{DOC_radiation})
\end{itemize}

\subsubsection{\label{DV}Deviations and counter measures}
Deviations and counter measures were not required for the completion of this deliverable.

\newpage
\section{Description of individual Dwarfs}
This section contains the description of the dwarfs presented in this deliverable including instructions to install and run them. References are listed at the end of each subsection.
\subsection{\label{DOC_multigrid}Multigrid preconditioned elliptic solver}
\includegraphics[page=6, scale=1.0, clip, trim=3.5cm 2.9cm 1cm 6.7cm]{multigrid.pdf}
\newpage
\newcount\pdfpageno
\pdfpageno=7
\loop
	\includegraphics[page=\pdfpageno, scale=1.0, clip, trim=3.5cm 2.55cm 1cm 2.8cm]{multigrid.pdf}
	\newpage
	\advance \pdfpageno 1
\ifnum \pdfpageno<21
\repeat
\setcounter{equation}{22}
\newpage
\begin{refsection}
\subsection{\label{DOC_MPDATA_unstructured}MPDATA for unstructured meshes}
\input{MPDATA_unstructured}
\subsubsection{References for unstructured MPDATA}
\printbibliography[heading=none]
\end{refsection}
\begin{refsection}
\subsection{\label{DOC_MPDATA_structured}MPDATA for structured meshes}
\input{MPDATA_structured}

\subsubsection{References for structured MPDATA}
\printbibliography[heading=none]
\end{refsection}
\begin{refsection}
\subsection{\label{DOC_radiation}ACRANEB2 radiation scheme}
\input{radiation}

\subsubsection{References for ACRANEB2}
\printbibliography[heading=none]
\end{refsection}

\section{\label{CON}Conclusions}
A second set of weather and climate prediction model sub-components (dwarfs) has been established, documented and made available for uptake and testing by the ESCAPE project partners. These new dwarfs extend the range of computational characteristics in terms of memory bandwidth, communication and computational cost. The new multigrid preconditioned elliptic solver poses new challenges in terms of next neighbour communication. The two MPDATA dwarfs allow an interesting insight into comparing the optimisation potential of structured and unstructured meshes. The radiation dwarf represents a very important part of the physics computation and forms together with the cloud microphysics dwarf from the first batch the second dwarf in the physics category.

The completion of this deliverable allows to progress with subsequent tasks of ESCAPE dealing with code adaptation and performance evaluation on the available hardware architectures. The work on porting these new dwarfs to GPUs and Xeon Phi processors as well as using domain specific languages has started in working packages 2 and 3. First results have been reported in D3.3.

The deliverable has been produced on time and its outcome has been disseminated to project partners through the proposed mechanisms of ESCAPE.

\backmatter



\cleardoublepage\pagestyle{empty}
\includepdf[pages={6,7}]{ESCAPE-D1_2-frontAndBack.pdf}

\end{document}

%% file: MPDATA_unstructured.tex
\subsubsection{Scope}
MPDATA stands for multidimensional positive definite advection transport algorithm. We have two versions of this dwarf: one for structured meshes and one for unstructured meshes. The version for structured meshes is used in the COSMO-EULAG model. Measurements have shown that depending on the number of processors used about 25\% to 34\% of the entire runtime of the model is spent in the MPDATA dwarf. The unstructured version described in this document is used in a newly developed finite volume dynamical core at ECMWF called FVM. We expect similar importance of this dwarf in FVM like in COSMO-EULAG.

\subsubsection{Objectives}
As described before the MPDATA algorithm takes a significant amount of the runtime of FVM. For this reason the key objectives of this dwarf are
\begin{itemize}
	\item the optimisation of the runtime per time-step of the MPDATA algorithm and
	\item exploration of new hardware architectures like GPUs and Xeon Phi processors to accelerate its computation
\end{itemize}
The dwarf offers an isolated prototype for advection on the sphere using an unstructured mesh with the MPDATA algorithm to facilitate work on optimising its performance and exploring new processors.

\subsubsection{Definition}
\paragraph{Fundamental concept}
The fundamental idea behind the MPDATA algorithm can be illustrated for the 1D advection equation of a field $\psi(x,t)$ with a constant advection velocity $v>0$:
\begin{align}
\label{eq:pde}
	\frac{\partial\psi}{\partial t}+\frac{\partial}{\partial x}\left(v\,\psi\right)=0.
\end{align}
The analytical solution of this equation is given by
\begin{align}
	\psi(x,t)=\psi_0(x-v\,t).
\end{align}
with the initial distribution $\psi_0(x)=\psi(x,t_0)$ at some initial time $t_0$.
To solve (\ref{eq:pde}) numerically we discretise time and space by introducing grid points (nodes) $x_i$ and time steps $t_n$. The simplest method to solve (\ref{eq:pde}) is the so called upwind scheme in which we approximate the time derivative by a finite difference forward in time and the spatial derivative by a left sided finite difference (for $v>0$). This gives us
\begin{align}
\label{eq:upwind}
	\psi_i^{n+1}=\psi_i^n-\frac{\delta t\,v}{\delta x}(\psi_i^n-\psi_{i-1}^n)
\end{align}
To make this scheme more accurate we use Taylor expansion up to second order in $\delta x$ and $\delta t$:
\begin{align}
	\psi_i^{n+1}&\approx\psi_i^n+\delta t\frac{\partial\psi}{\partial t}+\frac{1}{2}\delta t^2\frac{\partial^2\psi}{\partial t^2}\\
	\psi_{i-1}^n&\approx\psi_i^n-\delta x\frac{\partial\psi}{\partial x}+\frac{1}{2}\delta x^2\frac{\partial^2\psi}{\partial x^2}
\end{align}
Introducing these Taylor expansions into the upwind scheme (\ref{eq:upwind}) leads to
\begin{align}
\label{eq:taylorstep1}
	\psi_i^n+\delta t\frac{\partial\psi}{\partial t}+\frac{1}{2}\delta t^2\frac{\partial^2\psi}{\partial t^2}=\psi_i^n-\frac{\delta t\,v}{\delta x}\left[\delta x\frac{\partial\psi}{\partial x}-\frac{1}{2}\delta x^2\frac{\partial^2\psi}{\partial x^2}\right]
\end{align}
We can eliminate the second derivative in time by using the PDE (\ref{eq:pde}):
\begin{align}
	\frac{\partial^2\psi}{\partial t^2}\overset{(\ref{eq:pde})}{=}-v\frac{\partial^2\psi}{\partial t\partial x}\overset{(\ref{eq:pde})}{=}v^2\frac{\partial^2\psi}{\partial x^2}
\end{align}
Using this in (\ref{eq:taylorstep1}) and rearranging the terms leads to the following equation
\begin{align}
	\frac{\partial\psi}{\partial t}=-\frac{\partial}{\partial x}(v\,\psi)-\frac{\partial}{\partial x}\left[-\frac{v\,\delta x}{2}\frac{\partial\psi}{\partial x}+\frac{v\,\delta t}{2}\frac{\partial}{\partial x}(v\,\psi)\right].
\end{align}
We can interpret the last term in the square brackets for $\psi\neq0$ as an anti-diffusive flux. This can be illustrated by defining an anti-diffusive pseudo-velocity $v_\text{ad}$ which allows to write the equations which MPDATA solves as
\begin{align}
\label{eq:pde2}
	\frac{\partial\psi}{\partial t}&=-\frac{\partial}{\partial x}(v\,\psi)-\frac{\partial}{\partial x}(v_\text{ad}\,\psi).\\
	\label{eq:pseudovel1}
	v_\text{ad}&=-\frac{v\,\delta x}{2\psi}\frac{\partial\psi}{\partial x}+\frac{v\,\delta t}{2\psi}\frac{\partial}{\partial x}(v\,\psi).
\end{align} 

\paragraph{Generalisation for the Sphere}
The unstructured MPDATA dwarf solves the following more general equation
\begin{align}
	\frac{\partial(G\,\psi)}{\partial t}+\frac{\partial}{\partial x}\left(v\,\psi\right)=0,
\end{align}
where the symbol $G$ describes geometric and physical terms (Jacobian of coordinate transformations and fluid density) which are also included here in a modified advection velocity $v$. We use in this dwarf the so called infinite-gauge version of MPDATA which allows the field $\psi$ to change its sign by essentially removing $\psi$ from the denominator in (\ref{eq:pseudovel1}). This gives us together with the introduction of $G$ the following expression for the pseudo-velocity $v_\text{ad}$:
\begin{align}
\label{eq:pseudovel}
	v_\text{ad}=-\frac{1}{2}|v|\,\frac{\partial\psi}{\partial x}\,\delta x+\frac{1}{2}v\,\delta t\left[\frac{\partial}{\partial x}\left(v\,\psi\right)+\psi\,\frac{\partial G}{\partial t}\right].
\end{align}
Descretising (\ref{eq:pde2}) with forward difference in time gives
\begin{align}
		\psi_i^{n+1}=\psi_i^n-\delta t\frac{\partial}{\partial x}(v\,\psi)-\delta t\frac{\partial}{\partial x}(v_\text{ad}\,\psi).
\end{align}
The spatial derivatives are discretised with upwind fluxes which results in three dimensions in a sum of the upwind fluxes over all interfaces between neighbouring grid cells. The derivation of the multidimensional case and more details about its implementation are described in \cite{Kuehnlein2016}. A pre-print of this paper can be found in doc/MPDATA-JCP2016.pdf in the git-repo.

\paragraph{Implementation}
To illustrate the implementation of MPDATA we indicate the order of the different functions by the horizontal coloured bars beneath the following two equations. These equations are only meant to be a basic illustration of the code. Differently from these equations the full code is three dimensional.
\begin{center}
\resizebox{9cm}{5cm}{\includegraphics{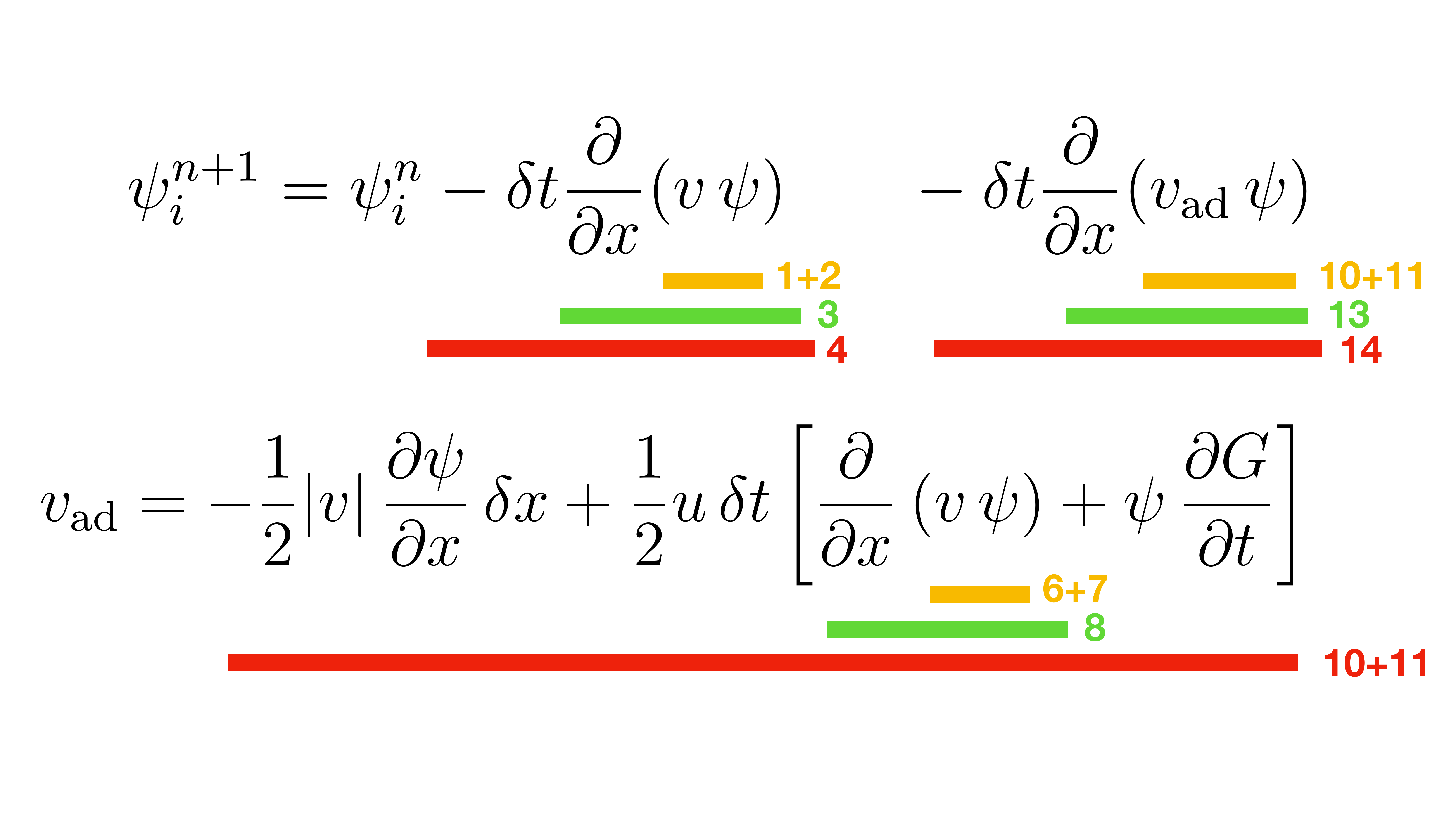}}
\end{center}
The numbers next to each bar stand for one of the functions of the dwarf as given by the following pseudo-code:
\begin{itemize}
	\item[] main program: loop over all timesteps. In each step:
    \item[]\textbf{1. compute\_upwind\_flux}:
        \subitem for all edges: for all levels:
            \subsubitem compute the upwind flux for that edge
    \item[]\textbf{2. compute\_upwind\_fluz}:
        \subitem for all nodes: for all levels:
            \subsubitem compute the upwind flux between vertically neighbouring grid cells
    \item[]\textbf{3. compute\_fluxzdiv}: 
        \subitem for all nodes: for all levels:
            \subsubitem compute sum over all horizontal and vertical fluxes weighted with
            \subsubitem the surface of the corresponding face
    \item[]\textbf{4. advance\_solution}:
        \subitem for all nodes: for all levels:
            \subsubitem add computed sum to the advected field pD multiplied with the
            \subsubitem timestep and divided through the density
    \item[]\textbf{5. rhofac\_correction}:
        \subitem for all nodes: for all levels:
            \subsubitem compute product of density and advected field
    \item[]\textbf{6. compute\_centered\_flux}:
        \subitem for all edges: for all levels:
            \subsubitem compute centred (averaged) flux for that edge
    \item[]\textbf{7. compute\_centered\_fluz}:
        \subitem for all nodes: for all levels:
            \subsubitem compute centred flux between vertically neighbouring grid cells
    \item[]\textbf{8. compute\_fluxzdiv}: 
        \subitem for all nodes: for all levels: 
            \subsubitem compute sum over all horizontal and vertical centred fluxes
            \subsubitem weighted with the surface of the corresponding face
    \item[]\textbf{9. halo\_exchange}:
        \subitem communicate results between neighbouring processors
    \item[]\textbf{10. compute\_pseudovel\_xy}:
        \subitem for all edges: for all levels:
            \subsubitem compute first term in the pseudo-velocity for that edge
        \subitem for all edges: for all levels:
            \subsubitem compute second term in the pseudo-velocity and add to first term
    \item[]\textbf{11. compute\_pseudovel\_z}:
        \subitem for all nodes: for all levels:
            \subsubitem compute first term in the pseudo-velocity between vertically
            \subsubitem neighbouring grid cells
        \subitem for all nodes: for all levels:
            \subsubitem compute second term in the pseudo-velocity and add to first term
            \subsubitem for vertically neighbouring grid cells
    \item[]\textbf{12. limit\_flux}
    \item[]\textbf{13. compute\_fluxzdiv}:
        \subitem for all nodes: for all levels:
            \subsubitem compute sum over all horizontal and vertical antidiffusive fluxes
            \subsubitem weighted with the surface of the corresponding face
    \item[]\textbf{14. advance\_solution}:
        \subitem for all nodes: for all levels:
            \subsubitem add computed sum to the advected field pD multiplied with the
            \subsubitem timestep and divided through the density
    \item[]\textbf{15. halo\_exchange}:
        \subitem communicate results between neighbouring processors
\end{itemize}
\subsubsection{Dwarf usage and testing}
In this section we describe how to download and install 
the dwarf along with all its dependencies, and we show 
how to run it for a simple test case.

Note that the MPDATA dwarf for unstructured meshes is implemented using \textit{Atlas}, 
the ECMWF software framework that supports flexible 
data-structures for NWP. The dwarf is written in Fortran 2003. Extensions to C++ can 
be envisioned if necessary and they can be implemented 
using \textit{Atlas}.

\paragraph{Download and installation}
The first step is to download and install the dwarf along 
with all its dependencies. With this purpose, it is possible 
to use the script provided under the ESCAPE software collaboration 
platform:\\
\url{https://git.ecmwf.int/projects/ESCAPE}.

Here you can find a repository called \inlsh{escape}.
You need to download it. There are two options to do this. One option is to use ssh. For this option you need to add an ssh key to your bitbucket account at \url{https://git.ecmwf.int/plugins/servlet/ssh/account/keys}. The link "SSH keys" on this website gives you instructions on how to generate the ssh key and add them to your account. Once this is done you should first create a 
folder named, for instance, ESCAPE, enter into it 
and subsequently download the repository by using the following the steps below:
\begin{lstlisting}[style=BashStyle]
mkdir ESCAPE
cd ESCAPE/
git clone ssh://git@git.ecmwf.int/escape/escape.git
\end{lstlisting}
The other option to download the repo is by using https instead of ssh. Instead of the git command above you then need to use 
\begin{lstlisting}[style=BashStyle]
git clone https://<username>@git.ecmwf.int/scm/escape/escape.git
\end{lstlisting}
where <username> needs to be replace by your bitbucket username.

Once the repository is downloaded into the \inlsh{ESCAPE} folder 
just created, you should find a new folder called \inlsh{escape}. 
The folder contains a sub-folder called \inlsh{bin} that has the 
python/bash script (called \inlsh{escape}) that needs to be 
run for downloading and installing the dwarf and its dependencies. 
To see the various options provided by the script you can type:
\begin{lstlisting}[style=BashStyle]
./escape/bin/escape -h
\end{lstlisting}
To download the dwarf you need to run 
the following command:
\begin{lstlisting}[style=BashStyle]
./escape/bin/escape checkout dwarf-D-advection-MPDATA \ 
--ssh
\end{lstlisting}
To use https you need to replace --ssh with --user <username>. The commands above automatically check out the \inlsh{develop}
version of the dwarf. If you want to download a specific branch 
of this dwarf, you can do so by typing:
\begin{lstlisting}[style=BashStyle]
./escape/bin/escape checkout dwarf-D-advection-MPDATA --user <username> \
--version <branch-name>
\end{lstlisting}
You should now have a folder called 
\inlsh{dwarf-D-advection-MPDATA}.

At this stage it is possible to install the dwarf 
and all its dependencies. This can be done in two 
different ways. The first way is to compile and 
install each dependency and the dwarf separately:
\begin{lstlisting}[style=BashStyle]
./escape/bin/escape generate-install dwarf-D-advection-MPDATA
\end{lstlisting}
The command above will generate a script 
called \inlsh{install-dwarf-D-advection-MPDATA} 
that can be run by typing:
\begin{lstlisting}[style=BashStyle]
./install-dwarf-D-advection-MPDATA
\end{lstlisting}
This last step will build and install the dwarf 
along with all its dependencies in the following 
paths:
\begin{lstlisting}[style=BashStyle]
dwarf-D-advection-MPDATA/builds/
dwarf-D-advection-MPDATA/install/
\end{lstlisting}

The second way is to create a bundle that compiles 
and installs all the dependencies together:
\begin{lstlisting}[style=BashStyle]
./escape/bin/escape generate-bundle dwarf-D-advection-MPDATA
\end{lstlisting}
This command will create an infrastructure to avoid
compiling the single third-party libraries individually
when some modifications are applied locally to one of 
them. To complete the compilation and installation process, 
after having run the above command for the bundle, simply 
follow the instructions on the terminal.

In the commands above that generate the installation 
file, you can specify several other optional parameters. 
To see all these options and how to use them you 
can type the following command:
\begin{lstlisting}[style=BashStyle]
./escape generate-install -h
./escape generate-bundle -h
\end{lstlisting}

\paragraph{Testing}
You should now verify that the dwarf works as expected.
For this purpose, we created a testing framework that
allows us to verify that the main features of the dwarf 
are working correctly.

In particular, for each sub-dwarf we provide various 
regression tests in order to allow the results to be 
consistent when the underlying algorithms are modified, 
and to test additional features or different hardware. 
The regression tests can be found in the folder \inlsh{test} 
that is located in each sub-dwarf folder. 
For each sub-dwarf we also provide scripts running the 
code on different architectures, e.g. the Cray HPC at 
ECMWF, that can be found in the folder \inlsh{run-scripts} 
located in each sub-dwarf folder. 
\begin{tipbox}
We encourage partners who are testing different architectures 
to add and/or modify the scripts!
\end{tipbox}

To run this verification, you should run the following 
command:
\begin{lstlisting}[style=BashStyle]
ctest -j<number-of-tasks>
\end{lstlisting}
from inside the \inlsh{builds/dwarf-D-advection-MPDATA}
folder.
\begin{warningbox}
We strongly advise you to verify via ctest that 
the main functionalities of the dwarf are working 
properly any time you apply modifications to the 
code. Updates that do not pass the tests cannot 
be merged. 
In addition, if you add a new feature to the dwarf,
this should be supported by a test if the existing
testing framework is not already able to verify its
functionality.
\end{warningbox}
For instructions on how to run the executables 
see the next section.
\begin{tipbox}
The tests rely on the tolerance specified in the config-files. This tolerance has been chosen in such a way that single and double precision simulations pass the test if the build type is set to Bit. This has been tested successfully at ECMWF with GFortran, Intel (version 16.0.3) and Cray compilers. However, the tests might fail with some compilers (at ECMWF with the Cray compiler) when using build type Release. If tests fail on your platform please make sure that you use build type Bit!
\end{tipbox}

\subsubsection{Run the Dwarf}
If you want to run the dwarf in your local machine, 
you could do so by using the executable files inside 
\begin{lstlisting}[style=BashStyle] 
dwarf-D-advection-MPDATA/install/dwarf-D-advection-MPDATA/bin/
\end{lstlisting}
In this section we assume that the executable has been generate with the generate-install option of the escape script. If you used the generate-bundle option you should replace \inlsh{install/dwarf-D-advection-MPDATA} with \inlsh{builds/bundle/bin}.

The executables need the specification 
of a configuration file. The configuration files can be found at
\begin{lstlisting}[style=BashStyle] 
dwarf-D-advection-MPDATA/sources/dwarf-D-advection-MPDATA/config-files/
\end{lstlisting}
The executable can be run as follows:
\begin{lstlisting}[style=BashStyle] 
dwarf-D-advection-MPDATA/install/dwarf-D-advection-MPDATA/\
bin/dwarf_D_advection_MPDATA --config \
dwarf-D-advection-MPDATA-solidBody-pole-O32.json
\end{lstlisting}
where, if the \inlsh{.json} file is not in the current directory, 
you can specify its path after \inlsh{-\,-config}. We used \inlsh{dwarf-D-advection-MPDATA-solidBody-pole-O32.json} in the last command as an example for one of the configuration files. This configuration file contains the parameters for a solid body rotation over the poles. There are also configuration files for a rotation along the equator (\inlsh{dwarf-D-advection-MPDATA-solidBody-equator-O32.json}) as well as a rotation along a direction in between those two (\inlsh{dwarf-D-advection-MPDATA-solidBody-diagonal-O32.json}). The configuration files write by default output-files every 100 timesteps which can be plotted with gmsh (\url{http://gmsh.info}). There are also copies of these configuration files ending with \inlsh{-noOut} where the parameter iout is set to 0 which does not create any output. These \inlsh{-noOut} configuration files should be used for performance measurements.

If you instead want to run the dwarf on an HPC machine 
available to the ESCAPE partners, you can automatically 
generate the job submission script with the \inlsh{escape} 
file. 

More specifically, if you run the following command:
\begin{lstlisting}[style=BashStyle]
./escape generate-run -c \
"dwarf-D-advection-MPDATA/install/dwarf-D-advection-MPDATA/\
bin/dwarf_D_advection_MPDATA --config \
dwarf-D-advection-MPDATA-solidBody-pole-O32.json"
\end{lstlisting}
This allows the code to generate the submission script      
for the given HPC machine you are targeting without submitting
the actual job. The command above will in fact simply generate
an \inlsh{escape.job} file in the current folder. This can
successively be submitted via \inlsh{qsub} on the HPC machine
you want to run the simulation on.

In the above command you can specify several other optional 
parameters, such as wall-time, number of tasks, number of 
threads, etc. To see all these options and how to set them 
up you can type the following command:
\begin{lstlisting}[style=BashStyle]
./escape generate-run -h
\end{lstlisting}
The following subsections describe how the precision of the computation can be selected and the data generated by the simulation in the log- and output-files.
\paragraph{Switching between single and double precision}
The dwarf has been tested with single as well as double precision. The error tolerance in the configuration files has been chosen in such a way that all compilers should pass the test with both single and double precision if build type Bit is used. To switch between single and double precision please adjust the line \inlsh{integer, parameter, public :: wp = sp} inside the file \inlsh{dwarf\_D\_advection\_MPDATA\_auxiliary\_module.F90}. Setting \inlsh{wp = dp} in this line uses double precision for the entire simulation whereas \inlsh{wp = sp} uses single precision.
\paragraph{Log and output data}
\subparagraph{Error measures}
In each output-step as defined by the parameter \inlsh{iout} in the configuration file the code measures the accuracy of the result by comparing the numerical result with an analytical result. For better comparison with the literature \cite{Smolarkiewicz1990} a number of different error measures are computed:
\begin{itemize}
	\item EMIN: error of the minimum value of the advected tracer
	\item EMAX: error of the maximum value of the advected tracer
	\item ERR0 (L2-error): root-mean-square error of the solution
	\item ERR1: normalised error of the mean field
	\item ERR2: variance of the field
	\item Linf-error: maximum of the error
\end{itemize}
The most important error measures are the L2- and Linf-error. Only these two are considered when comparing the result with the reference result stored in the configuration file.
\subparagraph{Wallclock-time of the timeloop}
The code uses the Fortran function \inlsh{system\_clock} to measure the time between starting and finishing the timeloop. This measurement should be used when comparing different computer architectures. For accelerators please also include a measurement that includes the process of copying the data to the device. Please report your measurements (no matter if they are good or bad) on \url{https://confluence.ecmwf.int/display/ESCAPE/Dwarf+-+D+-+advection+-+MPDATA} by following the example under the section "Performance Measurements" of that website.

%% file: MPDATA_structured.tex
\subsubsection{Scope}
MPDATA stands for multidimensional positive definite advection transport algorithm. We have two versions of this dwarf: one for structured meshes and one for unstructured meshes. The version for structured meshes described in this document is used in the COSMO-EULAG model for regional NWP. Measurements have shown that depending on the number of processors used about 25\% to 34\% of the entire runtime of the model is spent in the MPDATA dwarf. The unstructured version is used in a newly developed finite volume dynamical core at ECMWF called FVM. We expect similar importance of this dwarf in FVM like in COSMO-EULAG.

\subsubsection{Objectives}
Since this dwarf may significant share of the runtime of the weather model (as the number of passive tracers e.g. chemical species grows, the share can be very large), it is important that we optimise it as much as possible and explore the use of GPUs and Xeon Phi processors to accelerate its computation. This is the key objective behind this dwarf.

\subsubsection{Definition}
\paragraph{Fundamental concept}
The fundamental idea behind the MPDATA algorithm can be illustrated for the 1D advection equation of a field $\psi(x,t)$ with a constant advection velocity $v>0$:
\begin{align}
\label{eq:pde}
	\frac{\partial\psi}{\partial t}+\frac{\partial}{\partial x}\left(v\,\psi\right)=0.
\end{align}
The analytical solution of this equation is given by
\begin{align}
	\psi(x,t)=\psi_0(x-v\,t).
\end{align}
with the initial distribution $\psi_0(x)=\psi(x,t_0)$ at some initial time $t_0$.
To solve (\ref{eq:pde}) numerically we discretise time and space by introducing grid points (nodes) $x_i$ and time steps $t_n$. The simplest method to solve (\ref{eq:pde}) is the so called upwind scheme in which we approximate the time derivative by a finite difference forward in time and the spatial derivative by a left sided finite difference (for $v>0$). This gives us
\begin{align}
\label{eq:upwind}
	\psi_i^{n+1}=\psi_i^n-\frac{\delta t\,v}{\delta x}(\psi_i^n-\psi_{i-1}^n)
\end{align}
To make this scheme more accurate we use Taylor expansion up to second order in $\delta x$ and $\delta t$:
\begin{align}
	\psi_i^{n+1}&\approx\psi_i^n+\delta t\frac{\partial\psi}{\partial t}+\frac{1}{2}\delta t^2\frac{\partial^2\psi}{\partial t^2}\\
	\psi_{i-1}^n&\approx\psi_i^n-\delta x\frac{\partial\psi}{\partial x}+\frac{1}{2}\delta x^2\frac{\partial^2\psi}{\partial x^2}
\end{align}
Introducing these Taylor expansions into the upwind scheme (\ref{eq:upwind}) leads to
\begin{align}
\label{eq:taylorstep1}
	\psi_i^n+\delta t\frac{\partial\psi}{\partial t}+\frac{1}{2}\delta t^2\frac{\partial^2\psi}{\partial t^2}=\psi_i^n-\frac{\delta t\,v}{\delta x}\left[\delta x\frac{\partial\psi}{\partial x}-\frac{1}{2}\delta x^2\frac{\partial^2\psi}{\partial x^2}\right]
\end{align}
We can eliminate the second derivative in time by using the PDE (\ref{eq:pde}):
\begin{align}
	\frac{\partial^2\psi}{\partial t^2}\overset{(\ref{eq:pde})}{=}-v\frac{\partial^2\psi}{\partial t\partial x}\overset{(\ref{eq:pde})}{=}v^2\frac{\partial^2\psi}{\partial x^2}
\end{align}
Using this in (\ref{eq:taylorstep1}) and rearranging the terms leads to the following equation
\begin{align}
 \label{eq:upwindplusantidiff}
	\frac{\partial\psi}{\partial t}=-\frac{\partial}{\partial x}(v\,\psi)-\frac{\partial}{\partial x}\left[-\frac{v\,\delta x}{2}\frac{\partial\psi}{\partial x}+\frac{v\,\delta t}{2}\frac{\partial}{\partial x}(v\,\psi)\right].
\end{align}
We can interpret the last term in the square brackets for $\psi\neq0$ as an anti-diffusive flux. This can be illustrated by defining an anti-diffusive pseudo-velocity $v_\text{ad}$ which allows to write the equations which MPDATA solves as
\begin{align}
\label{eq:pde2}
	\frac{\partial\psi}{\partial t}&=-\frac{\partial}{\partial x}(v\,\psi)-\frac{\partial}{\partial x}(v_\text{ad}\,\psi).\\
	\label{eq:pseudovel1}
	v_\text{ad}&=-\frac{v\,\delta x}{2\psi}\frac{\partial\psi}{\partial x}+\frac{v\,\delta t}{2\psi}\frac{\partial}{\partial x}(v\,\psi).
\end{align} 
For the details of multidimensional definition of antidiffusive velocity, please refer to \cite{SMOLARKIEWICZ1998459}.
\paragraph{Limiters for antidiffusive fluxes}
The discussion on the FCT option of MPDATA can be found in section 3.2 of \cite{smolarkiewicz1990multidimensional}.  It begins with the search for the limiters (spatial indices were dropped for convenience):
\begin{align}
\label{eg:limiters}
\psi^{MAX}=max\left(\psi^{ant},\psi^{ant}_{neighbours},max(\psi^{in},\psi^{in}_{neighbours})\right)
\end{align}
where $\psi^{ant}$ is the first order approximation of $\psi$ at time $n+1$ (after the first upwind pass) and $\psi^{in}$ is the tracer field that enters the MPDATA procedure. The subscript $_{neighbours}$ describes set of values of $\psi$ at the total of $6$ nearest neighbouring points on A grid, that is $i+1,j,k$,$i,j+1,k$ and so on. Next, limiting ratios are evaluated such that:
\begin{align}
\label{eq:betas}
cp&=&\frac{\psi^{MAX}-\psi^{ant}}{\sum_{I=1}^{3} \left(-min(0,v^{ad}_{I,left})+max(0,v^{ad}_{I,right})\right)+ep}\\
cn&=&\frac{\psi^{ant}-\psi^{MIN}}{\sum_{I=1}^{3} \left( max(0,v^{ad}_{I,left})-min(0,v^{ad}_{I,right})\right)+ep}
\end{align}
\paragraph{Generalisation for the Sphere}
The unstructured MPDATA dwarf solves the following more general equation
\begin{align}
	\frac{\partial(G\,\psi)}{\partial t}+\frac{\partial}{\partial x}\left(v\,\psi\right)=0,
\end{align}
where the symbol $G$ describes geometric and physical terms (Jacobian of coordinate transformations and fluid density). We use in this dwarf the so called infinite-gauge version of MPDATA which allows the field $\psi$ to change its sign by essentially removing $\psi$ from the denominator in (\ref{eq:pseudovel1}). This gives us together with the introduction of $G$ the following expression for the pseudo-velocity $v_\text{ad}$:
\begin{align}
\label{eq:pseudovel}
d	v_\text{ad}=-\frac{1}{2}|v|\,\frac{\partial\psi}{\partial x}\,\delta x+\frac{1}{2}u\,\delta t\left[\frac{\partial}{\partial x}\left(v\,\psi\right)+\psi\,\frac{\partial G}{\partial t}\right].
\end{align}
Discretising (\ref{eq:pde2}) with forward difference in time gives
\begin{align}
		\psi_i^{n+1}=\psi_i^n-\delta t\frac{\partial}{\partial x}(v\,\psi)-\delta t\frac{\partial}{\partial x}(v_\text{ad}\,\psi).
\end{align}
The spatial derivatives are discretised with upwind fluxes which results in three dimensions in a sum of the upwind fluxes over all interfaces between neighbouring grid cells. The derivation of the multidimensional case and more details about its implementation are described in \cite{Kuehnlein2016}. A pre-print of this paper can be found in doc/MPDATA-JCP2016.pdf in the git-repo.

\paragraph{Implementation}
The actual MPDATA algorithm is compactly coded in the $*mpdatadwarf.F$ and $*mpdatadwarf\_sphere.F90$ files. The code is organised around main stencil computations, whereas the formulation of boundary conditions is hidden in auxiliary routines that perform simple operations (copy, negation, zero) on the outer computational halo. MPI halo updates are performed by $update*$ routines that ultimately point to single halo update subroutine in $*mpi\_parallel.F90$.

The following pseudo-code exposes main stencil computations, while dismissing boundary conditions and halo updates that have minor effect on performance on typical number of cores. Typically, the algorithm is memory bound, so the motivation is to inform about the data access pattern characteristic to subsequent components of the MPDATA algorithm, as well as the range of computations sharing a single set of loops.  The example given considers gauge version of MPDATA for cartesian domain, the spherical and standard MPDATA bear similar structure.
\begin{itemize}
	\item[] main program driver: loop over all timesteps. In each step (loops over i,j,k dropped for compactness):
    \item[]\textbf{Evaluate upwind algorithm (i.e. fully 3D first order forward-in-time advection algorithm}:
    Here, $x\_in$ is the tracer input, $xant$ is an auxiliary variable corresponding to the tracer after first upwind pass, $rhr$ is the ratio of densities from the two subsequent timesteps (here equal unity, but preserved for  exact reproduction of weather solver memory bandwidth requirements), $hi$ is the inverse of density. Note that donor statement computes net flux through given finite volume boundary located at C-grid (where u1,u2 and u3 are defined).  This step refers to term $-\frac{\partial}{\partial x} (v\psi)$ of the eq. \ref{eq:pde2}.
    \begin{lstlisting}
!donor(y1,y2,a) = max(0.,a)*y1 - (-min(0.,a)*y2)
f1ijkp=donor(x_in(i  ,j,k),x_in(i+1,j,k),u1(i+1,j,  k  ))
f1ijk =donor(x_in(i-1,j,k),x_in(i  ,j,k),u1(i  ,j  ,k  ))
f2ijkp=donor(x_in(i,j  ,k),x_in(i,j+1,k),u2(i  ,j+1,k  ))
f2ijk =...;f3ijkp=...;f3ijk = ... 
xant(i,j,k)=rhr(i,j,k)*                           
          (x_in(i,j,k)-
                       ( f1ijkp-f1ijk             
                        +f2ijkp-f2ijk             
                        +f3ijkp-f3ijk )*hi(i,j,k))
    \end{lstlisting}

    \item[]\textbf{Evaluate antidiffusive velocities v1,v2,v3}:
    These are pseudovelocities that will be used in the second upwind iteration that corrects the excessive diffusion of the first-order upwind scheme. This is the most complex and computationally demanding component of MPDATA scheme. Formula for v1 is given below, whereas the formulas for v2 and v3 are very similar, only with permuted indices and transporting momenta components u1,u2,u3. Here h denotes the density. This step refers to evaluation of $v_{ad}$ in the term $-\frac{\partial}{\partial x} (v_{ad}\psi)$ of the  eq. \ref{eq:pde2}. 
     \begin{lstlisting}
!  vdyf(x1,x2,a,rinv)=(abs(a)-a**2*rinv)*rat2(x1,x2)
!   rat4(z0,z1,z2,z3)=(z3+z2-z1-z0)*.25

hmx= 1./(0.5*(h(i-1,j,k)+h(i,j,k)))
v1(i,j,k)=vdyf(xant(i-1,j,k),xant(i,j,k),u1(i,j,k),hmx)    
          -0.125*u1(i  ,j,k)*hmx*                          
               ((u2(i-1,j  ,k) + u2(i-1,j+1,k)                 
                +u2(i  ,j+1,k) + u2(i  ,j  ,k))               
         *rat4(xant(i-1,j-1,k),xant(i  ,j-1,k),              
               xant(i-1,j+1,k),xant(i  ,j+1,k))                    
                   +                                                                   
                (u3(i-1,j,k  ) + u3(i-1,j,k+1)
                +u3(i  ,j,k+1) + u3(i  ,j,k  ))
         *rat4(xant(i-1,j,k-1),xant(i  ,j,k-1),
               xant(i-1,j,k+1),xant(i  ,j,k+1)))
                ...
v2(i,j,k)=vdyf(xant(i,j-1,k),xant(i,j,k),u2(i,j,k),hmy)    
                ...
v3(i,j,k)=vdyf(xant(i,j,k-1),xant(i,j,k),u3(i,j,k),hmz)    
     \end{lstlisting}
    \item[]\textbf{Evaluate local maxima/minima of given tracer for limiters}:
     The max/min operation is taken for in-place and neighbours in main directions, for subsequent use in flux limiting. Note that these are evaluated for timestep n and the first order approximation of tracer field at n+1. This is the implementation of the eq. \ref{eg:limiters}.
\begin{lstlisting}
 mxijk_o=max(x_in(i-1,j  ,k  ),                     
             x_in(i  ,j  ,k  ), x_in(i+1,j  ,k  ), 
             x_in(i  ,j-1,k  ), x_in(i  ,j+1,k  ),
             x_in(i  ,j  ,k-1), x_in(i  ,j  ,k+1))

 mxijk  =max(xant(i-1,j  ,k  ), xant(i  ,j  ,k  ),  
             xant(i+1,j  ,k  ),        mxijk_o,     
             xant(i  ,j-1,k  ), xant(i  ,j+1,k  ),  
             xant(i  ,j  ,k-1), xant(i  ,j  ,k+1))
 mnijk_o= ...                     
 mnijk  = ...
\end{lstlisting}
    \item[]\textbf{Evaluate limiting members for flux-corrected transport}: 
    Note that small number $ep$ is introduced to avoid division by zero. This is the implementation of the eq. \ref{eq:betas}.
\begin{lstlisting}
!pp= max(0.,y)
!pn=-min(0.,y)
cp(i,j,k)=(mxijk-xant(i,j,k))*h(i,j,k)/        
          (pn(v1(i+1,j  ,k  ))+pp(v1(i,j,k))     
          +pn(v2(i  ,j+1,k  ))+pp(v2(i,j,k))    
          +pn(v3(i  ,j  ,k+1))+pp(v3(i,j,k))+ep)

cn(i,j,k)=(xant(i,j,k)-mnijk)*h(i,j,k)/      
          (pp(v1(i+1,j  ,k  ))+pn(v1(i,j,k))     
          +pp(v2(i  ,j+1,k  ))+pn(v2(i,j,k))    
          +pp(v3(i  ,j  ,k+1))+pn(v3(i,j,k))+ep)
\end{lstlisting}
    \item[]\textbf{Evaluate corrective upwind iteration}:
    Here second upwind pass is applied, effectively acting as negative diffusion. The result is passed to the input/output variable $x\_in$. This step alludes to the evaluation of the term $-\frac{\partial}{\partial x} (v_{ad}\psi)$ of the eq. \ref{eq:pde2}.
\begin{lstlisting}
f1ijk=                                                       
      pp(v1(i,j,k  ))*min(1.,cp(   ,j,k),cn(i-1,j,k))
     -pn(v1(i,j,k  ))*min(1.,cp(i-1,j,k),cn(i  ,j,k))
f1ijkp=                                                   
      pp(v1(i+1,j,k))*min(1.,cp(i+1,j,k),cn(i  ,j,k)) 
     -pn(v1(i+1,j,k))*min(1.,cp(i  ,j,k),cn(i+1,j,k))
f2ijk=                                                     
      pp(v2(i,j,k  ))*min(1.,cp(i,j  ,k),cn(i,j-1,k)) 
     -pn(v2(i,j,k  ))*min(1.,cp(i,j-1,k),cn(i,j  ,k))
f2ijkp= ...
f3ijk = ...                                                     
f3ijkp= ...
x_in(i,j,k)=(xant(i,j,k)/rhr(i,j,k)-
                        ( f1ijkp-f1ijk              
                         +f2ijkp-f2ijk               
                         +f3ijkp-f3ijk )*hi(i,j,k))

\end{lstlisting}
\end{itemize}
\subsubsection{Dwarf usage and testing}
In this section we describe how to download and install 
the dwarf along with all its dependencies, and we show 
how to run it for a simple test case.

Note that the MPDATA dwarf for structured meshes has no external dependences. Therefore, besides the generic ESCAPE build strategy it allows for manual build with Makefile, respective Makefiles are provided in $manualbuild*$ directories.
The dwarf is written in Fortran 95/2003.

\paragraph{Download and installation}
The first step is to download and install the dwarf along 
with all its dependencies. With this purpose, it is possible 
to use the script provided under the ESCAPE software collaboration 
platform:\\
\url{https://git.ecmwf.int/projects/ESCAPE}.

Here you can find a repository called \inlsh{escape}.
You need to download it. There are two options to do this. One option is to use ssh. For this option you need to add an ssh key to your bitbucket account at \url{https://git.ecmwf.int/plugins/servlet/ssh/account/keys}. The link "SSH keys" on this website gives you instructions on how to generate the ssh key and add them to your account. Once this is done you should first create a 
folder named, for instance, ESCAPE, enter into it 
and subsequently download the repository by using the following the steps below:
\begin{lstlisting}[style=BashStyle]
mkdir ESCAPE
cd ESCAPE/
git clone ssh://git@git.ecmwf.int/escape/escape.git
\end{lstlisting}
The other option to download the repo is by using https instead of ssh. Instead of the git command above you then need to use 
\begin{lstlisting}[style=BashStyle]
git clone https://<username>@git.ecmwf.int/scm/escape/escape.git
\end{lstlisting}
where <username> needs to be replace by your bitbucket username.

Once the repository is downloaded into the \inlsh{ESCAPE} folder 
just created, you should find a new folder called \inlsh{escape}. 
The folder contains a sub-folder called \inlsh{bin} that has the 
python/bash script (called \inlsh{escape}) that needs to be 
run for downloading and installing the dwarf and its dependencies. 
To see the various options provided by the script you can type:
\begin{lstlisting}[style=BashStyle]
./escape/bin/escape -h
\end{lstlisting}
To download the dwarf you need to run 
the following command:
\begin{lstlisting}[style=BashStyle]
./escape/bin/escape checkout dwarf-D-advection-MPDATA-structured \ 
--ssh
\end{lstlisting}
To use https you need to replace --ssh with --user <username>. The commands above automatically check out the \inlsh{develop}
version of the dwarf. If you want to download a specific branch 
of this dwarf, you can do so by typing:
\begin{lstlisting}[style=BashStyle]
./escape/bin/escape checkout dwarf-D-advection-MPDATA-structured --user <username> \
--version <branch-name>
\end{lstlisting}
You should now have a folder called 
\inlsh{dwarf-D-advection-MPDATA-structured}.

In the above command, you can specify several other optional 
parameters. To see all these options and how to use them you 
can type the following command:
\begin{lstlisting}[style=BashStyle]
./escape checkout -h
\end{lstlisting}

At this stage it is possible to install the dwarf 
and all its dependencies. This can be done in two 
different ways. The first way is to compile and 
install each dependency and the dwarf separately:
\begin{lstlisting}[style=BashStyle]
./escape/bin/escape generate-install dwarf-D-advection-MPDATA-structured
\end{lstlisting}
The command above will generate a script 
called \inlsh{install-dwarf-D-advection-MPDATA} 
that can be run by typing:
\begin{lstlisting}[style=BashStyle]
./install-dwarf-D-advection-MPDATA-structured
\end{lstlisting}
This last step will build and install the dwarf 
along with all its dependencies in the following 
paths:
\begin{lstlisting}[style=BashStyle]
dwarf-D-advection-MPDATA-structured/builds/
dwarf-D-advection-MPDATA-structured/install/
\end{lstlisting}

The second way is to create a bundle that compiles 
and installs all the dependencies together:
\begin{lstlisting}[style=BashStyle]
./escape/bin/escape generate-bundle dwarf-D-advection-MPDATA-structured
\end{lstlisting}
This command will create an infrastructure to avoid
compiling the single third-party libraries individually
when some modifications are applied locally to one of 
them. To complete the compilation and installation process, 
after having run the above command for the bundle, simply 
follow the instructions on the terminal.

In the commands above that generate the installation 
file, you can specify several other optional parameters. 
To see all these options and how to use them you 
can type the following command:
\begin{lstlisting}[style=BashStyle]
./escape generate-install -h
./escape generate-bundle -h
\end{lstlisting}
\paragraph{Manual building}
This dwarf has no external dependencies, so the build process can be easily controlled by hand. For this purpose, makefiles for each subdwarf are provided in the dwarf directory in 
\begin{lstlisting}[style=BashStyle]
./sources/dwarf-D-advection-MPDATA-structured/manualbuild-gauge
./sources/dwarf-D-advection-MPDATA-structured/manualbuild-standard
./sources/dwarf-D-advection-MPDATA-structured/manualbuild-sphere
\end{lstlisting}
The makefiles expose the preprocessor definitions mentioned above and contain a set of bit reproducible, standard and advanced optimisation options for Cray, NAG, PGI, Intel and GNU compilers. For porting of the makefile on given machine it should be sufficient to properly point $FTN\_NOMPI$ and $FTN\_MPI$ variables to the actual system commands invoking Fortran compilers in serial and parallel configuration.

\paragraph{Testing}
Contrary to unstructured advection dwarf, the testing is implemented within the dwarf binary itself. The MPDATA gauge and MPDATA gauge-sphere subdwarfs provide measure of L2 "energy" conservation error norm, for the reference configurations of $59^3$ cubic cartesian box (discussed recently  in section 3.7 in \cite{gmd-8-1005-2015}, and $128x64xL$ lon-lat-L spherical grid, originally defined in \cite{smolar1991monotone} and recently referred to in section 3.8 of  \cite{gmd-8-1005-2015}. The cartesian test represents three-dimensional sphere revolving around tilted axis with constant angular velocity; the error is measured exactly after one revolution. Standard MPDATA subdwarf is not provided with the reference solution as the reference results were not published for this particular setup, although the L2 error is computed as in the gauge subdwarf. In turn, the spherical test bases on the 2D solid-body rotation on a spherical surface. The trajectory of solid-body is set so it starts from the equator (where the grid is the relatively coarse) and it travels along meridian towards the pole, where the grid spacing in longitudinal direction is minimal. For each subdwarf, evaluation of the error norms at the end of integration can be deactived using TESTING definition in cmake script.\\

The dwarfs also employs custom execution timers for the actual dwarf subroutine and halo update routines (metrics are inclusive). The table appearing at the of integration provides information on number of calls to a particular routine, average timer on all cores along with maximum and minimum time of execution for given MPI process. This measurement should be used when comparing different computer architectures. For accelerators please also include a measurement that includes the process of copying the data to the device. Please report your measurements (no matter if they are good or bad) on \url{https://confluence.ecmwf.int/display/ESCAPE/Dwarf+-+D+-+advection+-+MPDATA+-+structured} by following the example under the section "Performance Measurements" of that website. This feature can be switched off with the TIMERSCPU definition in cmake file. \\
The dwarf allows for testing with static and dynamic memory allocation. This reveals the added value of the predefined size of matrices and loops at compile time. Static memory allocation can be controlled with STATICMEM define in cmake script.

\subsubsection{Run the Dwarf}
If you want to run the dwarf in your local machine, 
you could do so by using the executable subdwarf files inside 
\begin{lstlisting}[style=BashStyle] 
dwarf-D-advection-MPDATA-structured/install/\
dwarf-D-advection-MPDATA-structured/bin/
\end{lstlisting}

If you instead want to run the dwarf on an HPC machine 
available to the ESCAPE partners, you can automatically 
generate the job submission script with the \inlsh{escape} 
file. 

More specifically,  the following command (example for gauge subdwarf):
\begin{lstlisting}[style=BashStyle]
./escape generate-run -c \
"dwarf-D-advection-MPDATA-structured/install/\
dwarf-D-advection-MPDATA-structured/bin/\
dwarf_D_advection_MPDATA_structured_gauge"
\end{lstlisting}
allows the code to generate the submission script      
for the given HPC machine you are targeting without submitting
the actual job. The command above will in fact simply generate
an \inlsh{escape.job} file in the current folder. This can
successively be submitted via \inlsh{qsub} on the HPC machine
you want to run the simulation on.

In the above command you can specify several other optional 
parameters, such as wall-time, number of tasks, number of 
threads, etc. To see all these options and how to set them 
up you can type the following command:
\begin{lstlisting}[style=BashStyle]
./escape generate-run -h
\end{lstlisting}
\begin{warningbox}
However, note that for execution on more than one core you need to manually modify the $*parameters*.F90$ file for the given subdwarf, where parameters $nprocx$,$nprocy$,$nprocz$ are defined. Given dwarf must be then recompiled, as for the static memory allocation the size of MPI subdomain implicates the allocation of matrices performed at compile time.
\end{warningbox}
The following subsections describe how the precision of the computation can be selected and the data generated by the simulation in the log- and output-files.
\paragraph{Switching between single and double precision}
The dwarf has been tested with single as well as double precision. To switch between single and double precision please adjust the line \inlsh{INTEGER, PARAMETER :: euwp = sp} inside the file \inlsh{dwarf\_D\_advection\_MPDATA\_structured\_precisions.F90}. Setting \inlsh{euwp = dp} in this line uses double precision for the entire simulation whereas \inlsh{euwp = sp} uses single precision.

%% file: radiation.tex
\subsubsection{Scope}
\label{sec:scope}
The radiation schemes in numerical weather prediction and climate 
models take up a considerable amount of the overall running time 
of these models. Here, ``radiation'' is implicitly taken to mean
\emph{electromagnetic radiation}. The heating due to absorption of 
shortwave (solar) and longwave (terrestrial heat) radiation is the 
initial driver of all atmospheric processes with the exception of 
volcanic events.

Neither the shortwave nor the longwave radiative transfer can be 
solved within a reasonable amount of time from basic principles. 
In addition to the 
spatial and temporal approximations that are necessary to make for 
all physical processes in atmospheric models, it is necessary 
to make approximations in the spectral dimension and the directional 
dimensions. Thus, it is not feasible to calculate the radiative 
transfer for each absorbing and emitting line of the atmospheric 
gasses; in stead a limited number of spectral bands are defined for 
which the radiative transfer is calculated. For shortwave irradiance 
the radiative transfer in most current models is only considered for 
the direct solar beam, upward diffuse irradiance and downward diffuse 
irradiance. Thus, the complex directional variability of shortwave 
irradiances is not considered. This is called the two-stream 
approximation. For the longwave irradiances the two-stream 
approxiamation is also used in most current models. 
To sum up, many approximations 
are currently made in order to calculate radiative transfer in weather 
and climate models, and yet these are very resource demanding. A 
better utilisation of the radiation schemes on current and future 
extremely parallelised multi-threaded CPUs and GPUs is in demand.

Given these many ways radiation schemes can be approximated, they can 
be, and have been, designed in various ways. For radiation schemes in 
medium to long range weather models, it makes sense to utilise 
more spectral bands to capture the complex shortwave radiative heating 
in the stratosphere while saving resources in the spatial and temporal 
dimensions by running the radiation scheme in coarser resolution and 
intermittently relative to the general model time stepping. For 
radiation schemes in short range convective permitting weather models, 
on the other hand, it makes sense to resolve the diabatic heating 
patterns caused by small scale clouds, and not to use resources to 
improve the accuracy of the stratospheric heating rates. Here we have 
chosen to work with the ACRANEB2 radiation scheme 
\cite{Masek_et_al_2016,Geleyn_et_al_2017}, which is made for 
short range weather models. This dwarf has many of the features of 
physics modules in weather and climate models in general. Thus, it 
includes frequent usage of transcendental 
functions, and complex loop and conditional structures. This makes it 
interesting also in a broader context.

The overarching scope of this dwarf can be summarised in the following 
questions:

\begin{enumerate}
  \item What is the potential of refactoring the dwarf to run optimally 
        on 2016 model Xeon, Xeon Phi processors and GPUs?
  \item Is it worthwhile refactoring a physics subroutine such as this?
  \item Should the refactoring be done universally for the different 
        hardware architectures investigated?
  \item Can a set of rules be made for proper programming optimised for 
        current hardware architectures?
\end{enumerate}

Detailed descriptions of the physics in ACRANEB2 have been made by 
Ma{\v{s}}ek et al.~(2016) \cite{Masek_et_al_2016} and Geleyn et 
al.~(2017) \cite{Geleyn_et_al_2017} for the shortwave and longwave 
radiation, respectively. The work package 2 (WP2) software adaptation 
of the ACRANEB2 dwarf, and the benchmarking and diagnostics 
results for different hardware architectures (WP3) are detailed in the 
report by Poulsen and Berg (2017) \cite{Poulsen_et_Berg_2017}.
In this WP1 deliverable the focus is on how the original 
radiation scheme subroutines have been adapted to become a dwarf, and 
the subsequent dwarf versions developed from the initial version. How 
the dwarf should be run and can be modified is also described.

\subsubsection{Objectives}
\label{sec:objectives}
The main objective of this deliverable is to port the underlying code 
to accelerator and multithreaded architecture environments, and to 
document how the dwarf is run. The dwarf is then ready to be 
software-refactored (WP2) and tested on selected hardware 
architectures (WP3). 

The objectives are:
\begin{itemize} 
\item to make a stand-alone dwarf version of ACRANEB2 for different 
hardware architectures,
\item to make dwarfs of particularly computationally intensive parts of
      ACRANEB2 -- if needed, 
\item to check the reproducibility of the dwarf output for given 
      input for different compilers and compiler options, 
\item to measure the time-to-solution provided by implementations 
of the spectral transform on different hardwares, and 
\item to make a method for assessing the energy-to-solution.
\end{itemize}
To achieve these goals we have made two prototypes of the 
ACRANEB2 dwarf. The first prototype \inlsh{dwarf-lonlev-0.24} is a 
stand-alone version that can be run with given input data, namelist 
input, and which provides output as the original subroutine. Relative 
to the original version of the subroutine, the loop structure has been 
changed so that loops over model levels are the innermost loops, rather
than the outermost loops. Timing statements have also been added and 
other minimal changes to ensure portability have been made. The second 
prototype \inlsh{dwarf-lonlev-0.9} has been optimised for running on 2016 
model Xeon Phi processors as documented by Poulsen and Berg (2017)
\cite{Poulsen_et_Berg_2017}.

Furthermore, three prototypes based on a particularly computationally 
intensive part of ACRANEB2---the `transt3' dwarf---have been made. The 
first of these\\ \inlsh{dwarf-transt3\_v0.1} has been optimised for 
running on 2016 model Xeon Phi processors. The second two of these 
\inlsh{dwarf-transt3\_gpu\_v0.2} and\\ 
\inlsh{dwarf-transt3\_gpu\_nproma\_v0.1}
have been optimised for running on 2016 model GPUs. Here, the third 
transt3 version has loops that are tailor-made for the specific GPU 
tested.

\subsubsection{Description of Dwarf prototypes}

The Dwarf-P-ACRABNEB2-radiation scheme calculates atmospheric heating 
rates and specific downward surface fluxes for both shortwave and 
longwave radiation. To make a stand-alone version that can be run and 
tested as a dwarf, we have made a simple Fortran program structure 
that reads input from namelist files and modules, and writes formatted 
and binary output. As mentioned in section 
\ref{sec:objectives} we have two prototypes of the ACRANEB2 
dwarf. After installing the dwarf with the escape script as described above these prototypes can be found in the respective folders

\verb|escape-acraneb2-dwarf/src/dwarf-lonlev-0.24| and \\
\verb|escape-acraneb2-dwarf/src/dwarf-lonlev-0.9|.

In the following subsections these folders will be referred to with the 
generic name \verb|<dwarf>/|.

\paragraph{Namelist input variables}
\label{sec:namelists}

Namelists make it possible to run different experiments without having 
to recompile the code. We have made two namelists

\verb|escape-acraneb2-dwarf/test/escape-acraneb2-input/dimensions.nam| 
and \\
\verb|escape-acraneb2-dwarf/test/escape-acraneb2-input/nam_radia_dwarf.nam|.

In \verb|dimensions.nam| the spatial dimensions of the input 
data for the dwarf are specified as the number of longitudes 
\verb|KLO|, latitudes \verb|KLA| and levels \verb|KLEV| for the 
rectangular model grid. As is the case with most weather and climate 
model subroutines, the three dimensional space is reduced to
two dimensions \verb|KLON| and \verb|KLEV|, where the former is the 
total number of horizontal elements, that is the product of \verb|KLO|
and \verb|KLA|. This is done since the radiative (and other physical) 
processes are calculated as occurring in independent columns in the 
model grid. By default \verb|KLON = 160000| or 400 by 400 and 
\verb|KLEV = 80|.

In the namelist \verb|nam_radia_dwarf.nam| several variables are 
defined from which input data for the acraneb2 subroutine can be 
calculated. The point of this setup is to be able to easily make 
realistic input data without having to specify this for each grid box. 
Some of the variables in 
\verb|nam_radia_dwarf.nam| are unused in the default version, but could be utilised as desired.

The atmospheric model level input variables needed by the acraneb2 
subroutine are temperature (T), specific humidity (q), cloud cover 
(“NEB”), ice cloud load (“ICE”), liquid cloud load (“LI”), aerosol 
profiles (6 types), gasses (important for radiation), half level, 
full level pressures and pressure thicknesses. Additionally, the 
surface variables needed are diffuse and direct albedo, emittance and 
surface temperature. The full acraneb2 input is calculated in the 
subroutines \verb|<dwarf>/src/ini_var_ideal.F90| and 
\verb|<dwarf>/src/ini_acraneb.F90|.  

In \verb|nam_radia_dwarf.nam| variables that reflect the namelist 
variables that control how ACRANEB2 is run in both the full 
ALARO-1 and HARMONIE-AROME NWP models can be found. These are 
listed below.

\begin{itemize}
  \item \verb|LRNUMX| is a logical variable that chooses maximum-random 
                      cloud overlap. If the variable is set to false, 
                      random cloud overlap is chosen. 
                      For this version of ACRANEB2, 
                      it is recommended to be set to true.
  \item \verb|LCLSATUR| is a logical variable that makes the cloud
                        optical coefficients depend on the liquid/ice 
                        water content, and the saturation effect 
                        depend on cloud layers above and below. This 
                        should be set to true.
  \item \verb|LVOIGT| and \verb|LVFULL| are logical variables that 
                      accounts for the effect of Doppler line 
                      broadening in the mesosphere 
                      \cite{Masek_et_al_2016}, which results 
                      in the Voigt line shape. For short range 
                      forecasting models with the highest model levels 
                      in the stratosphere, this effect can be ignored. 
                      These variables are set to true by default.
  \item \verb|LRAYLU| is a logical variable that activates computation 
                      of lunar surface fluxes. This variable is set 
                      to false.
  \item \verb|LRPROX| is a logical variable that switches on 
                      exact adjacent exchanges of thermal irradiances. 
                      This should be switched off to get a better 
                      inclusion of cloudiness (pers. comm. Ma{\v{s}}ek 
                      2013) and is therefore set to false.
  \item \verb|LRTPP| is a logical variable that switches on 
                     nonisothermal layer correction in adjacent 
                     exchanges. This is set to true by default.
  \item \verb|LRAYPL| is a logical variable that accounts for using 
                      a shorter horizontal array for shortwave 
                      computations when the model domain is only 
                      partially daylit. This option has been removed 
                      in the refactored dwarf-lonlev-0.9.
  \item \verb|NPHYREP| is an integer variable for code tests. For 
                       this dwarf it should always be set to 1.
  \item \verb|NSORAYFR| is an integer variable that controls the 
                        intermittency of the full solar radiation 
                        computations. If it is 1, the full radiation is 
                        run at each time step, if is 2 it is run at 
                        every other time step, and so forth. If it is 
                        set to -1, the full radiation is run once 
                        every hour. For the original version of 
                        ACRANEB2 the default value is -1. For the 
                        refactored dwarf-lonlev-0.9 this option cannot 
                        be used and is thus always 1.
  \item \verb|NTHRAYFR| is an integer variable that controls the 
                        intermittency of the full thermal radiation 
                        computations. If it is 1, the full radiation is 
                        run at each time step, if is 2 it is run at 
                        every other time step, and so forth. If it is 
                        set to -1, the full radiation is run once 
                        every hour. For the original version of 
                        ACRANEB2 the default value is -1. For the 
                        refactored dwarf-lonlev-0.9 this option cannot 
                        be used and is thus always 1.
  \item \verb|NRAUTOEV| is an integer variable that controls the 
                        intermittency of the full computations of 
                        bracketing weights relative to the full 
                        computations of the thermal radiation. In the 
                        original version of ACRANEB2 the default value 
                        is 3 or every three hours.  For the 
                        refactored dwarf-lonlev-0.9 this option cannot
                        be used and is thus always 1.
\end{itemize}

\paragraph{Modules}

The original acraneb2 subroutine reads a large amount of natural 
constants and coefficients from module files. For the dwarf we have 
gathered all of these in one module \verb|<dwarf>/src/yomrad.mod| with 
its variables being defined in the subroutine 
\verb|<dwarf>/src/ini_modules.f90|. The module \verb|parkind.f90| is 
also used, which is a standard module defining variable types.

The module \verb|<dwarf>/src/dmi_timer.f90| has been developed by 
Jacob Weismann Poulsen and Per Berg (DMI) in order to ensure that 
timing estimates of the full dwarfs and their parts can be 
performed accurately and with a minimal amount of uncertainty. This 
is included in all dwarf prototypes. Calls to the timer subroutine from 
this module are added around the call to the acraneb2 subroutine 
itself, so that the overhead of the surrounding initialisation, input 
and output statements are not included in the time-to-solution 
computations.

\paragraph{Wrapping dwarf structure}
\label{sec:wrap}

The wrapping structure to call and test the ACRANEB2 dwarf consists of 
the simple Fortran program \verb|<dwarf>/src/callmain.f90|, which calls 
the subroutine \verb|<dwarf>/src/main.f90| from which the dimensions 
namelist and the subroutine \verb|<dwarf>/src/radia_dwarf.f90| are 
called. The subroutine \\
\verb|<dwarf>/src/radia_dwarf.f90| contains the 
calls to the other module and initialisation subroutines described 
above, the calls to the input and output subroutines, and the call 
to \verb|<dwarf>/src/acraneb2.f90|---the subroutine of primary interest.

\paragraph{The initial ACRANEB2 dwarf}
\label{sec:initial}

\begin{figure}[htb!]
\centering
\includegraphics[width=0.65\textwidth]{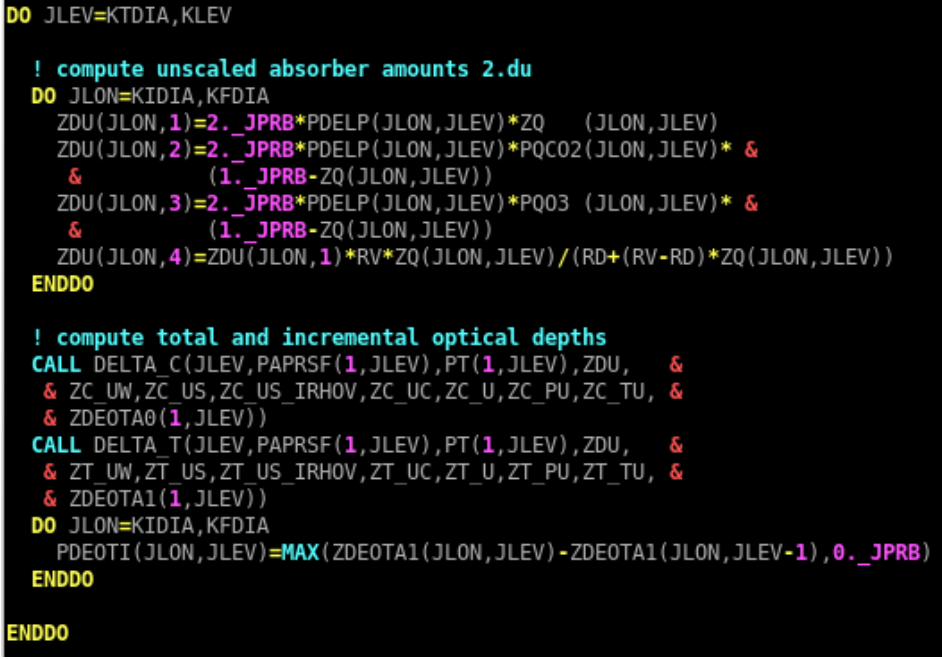}
\caption{Example from the original subroutine acraneb\_transt.f90 
         showing the original loop structure.}
\label{fig:transt_levlon_loop}
\end{figure}

\begin{figure}[htb!]
\centering
\includegraphics[width=0.65\textwidth]{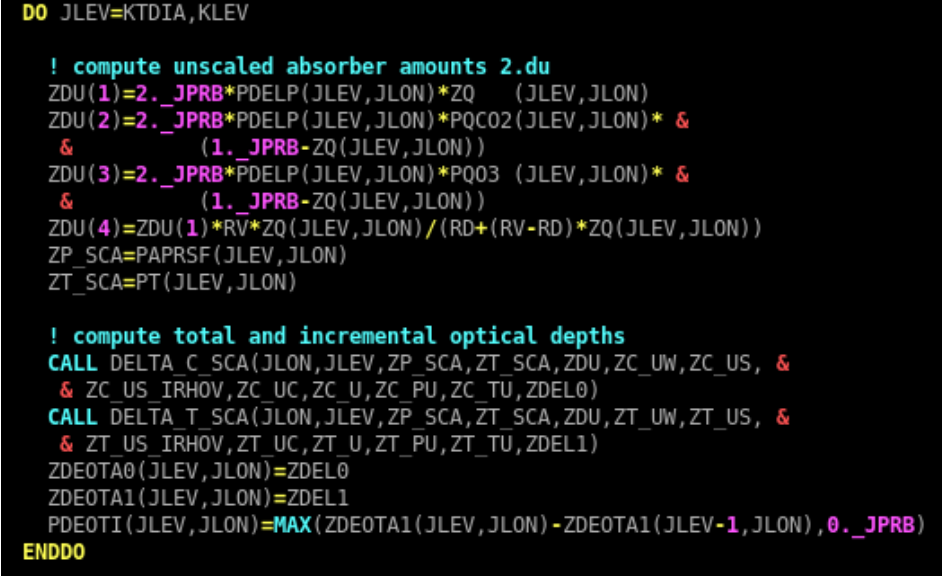}
\caption{Example from the initial dwarf version of the subroutine 
         acraneb\_transt.f90 showing the revised loop structure. This 
         is the same part of the subroutine as shown in
         Fig.~\ref{fig:transt_levlon_loop}. The JLON loop cannot 
         be seen here, as this is now moved to include a larger part of 
         of subroutine.}
\label{fig:transt_lonlev_loop}
\end{figure}

The initial prototype \inlsh{dwarf-lonlev-0.24} is modified relative to the 
original version of the subroutine: Minimal changes have been made in 
order to ensure portability, timing statements have been added, and 
the loop order has been changed so that the horizontal loops over 
model columns (JLON) are the outermost loops, while the loops over 
model levels (JLEV) are the innermost loops. The difference can be 
seen by comparing Fig.~\ref{fig:transt_levlon_loop} and 
Fig.~\ref{fig:transt_lonlev_loop}.
Given that the total number of model columns in a typical limited 
area model is of the orders 10$^5$--10$^6$ and the number of model 
levels is of the order of 10$^2$, the largest loops are now on the 
outside. When the loops are switched, the size of the local variables 
can be reduced to single columns and scalars, which reduces the 
stack memory load. For the exascale weather and climate models of the 
future the relative differences in the JLON vs the JLEV loops will be 
even larger.

The subroutine acraneb2 includes comments that labels the sections 
of the code. In sections I and II, initialisations,
 aerosol optical properties and preliminary calculations are made. 
Of primary importance is here 
the computation of the variables \verb|ICALS| and \verb|ICALT| that 
control the intermittency of the shortwave and longwave radiation 
computations, respectively. Here \verb|ICALS| depends on the time step 
\verb|KSTEP| and \verb|NSORAYFR| , while \verb|ICALT| depends on the 
\verb|KSTEP|, \verb|NTHRAYFR| and \verb|NRAUTOEV| as explained in 
section \ref{sec:namelists}. The default setup namelist 
\verb|nam_radia_dwarf.nam| enables \verb|ICALS = 0| and 
\verb|ICALT = 2|. In this part of the code the special case JLON loop 
indices \verb|IIDIA(IAUCR)| and \verb|IFDIA(IAUCR)| are also calculated.
These define the columns that are sunlit, in cases where the model 
domain is only partially sunlit, and replace the standard JLON loop 
indices \verb|KIDIA| and \verb|KFDIA|.

Next the gaseous optical properties are calculated in section III. This 
includes calls to the subroutines \verb|<dwarf>/src/acraneb_transs.f90| 
and \\ \verb|<dwarf>/src/acraneb_transt.f90| in which the (descending 
and ascending) optical depths in the shortwave and longwave 
spectral bands are calculated, respectively. The longwave gasseous 
optical depths are calculated both for the local temperatures and the 
temperature of the emitting body following the method of 
\cite{Ramanathan_et_Downey_1986}. The subroutine \verb|acraneb_transs| 
includes the subroutine \verb|delta_s_sca|. This has been modified 
from the original version \verb|delta_s|, where the added letters 
\verb|sca| are short for scalar and reflect the dimensional reduction 
of the variables that was enabled by rearranging the overall loop 
structure of the acraneb2 subroutine. Likewise \verb|acraneb_transt| 
includes calls to the subroutines \verb|delta_c_sca| and 
\verb|delta_t_sca| that are simplified versions of the original 
subroutines \verb|delta_c| and \verb|delta_t|.

After this, the cloud optical properties for the delta-two stream and 
adding radiative transfer computations are computed, mainly 
in the subroutine \\
\verb|<dwarf>/src/ac_cloud_model2.f90|. In section IV 
of the code cloud overlap effects are accounted for.

The longwave radiative transfer computations are performed in sections 
V--VIII. Here the subroutines \verb|<dwarf>/src/acraneb_coeft.f90|, \\
\verb|<dwarf>/src/acraneb_solvt.f90| and 
\verb|<dwarf>/src/acraneb_solvt3.f90| are \\ called. In the former 
matrix coefficients for the adding method linear system are computed, 
while this is solved by Gaussian elimination back-substitution in the 
two other subroutines.

The shortwave radiative transfer compuations are performed in section 
IX. Here the subroutines \verb|<dwarf>/src/acraneb_coefs.f90| and \\ 
\verb|<dwarf>/src/acraneb_solvs.f90| are called. In the former 
matrix coefficients for the adding method linear system are computed, 
while this is solved by Gaussian elimination back-substitution in the 
latter.

\paragraph{The refactored ACRANEB2 dwarf}
\label{sec:refactored}
~\\
\verb|escape-acraneb2-dwarf/src/dwarf-lonlev-0.9| is refactored for the 
optimal time-to-solution performance on the Xeon Phi targets (KNL-7210 
and KNL-7250). Regarding 
the overall structure, the main code has been gathered into the two 
modules \verb|acraneb2_m| and \verb|<dwarf>/src/acraneb3.f90|. Here 
the \verb|acraneb3| module contains all the subroutines that are called 
from the acraneb2 subroutine, while the \verb|acraneb_2m| module 
contains the primary subroutine acraneb2 and the added subroutines
\verb|acraneb2_heap| and \verb|acraneb2_numainit|. These two subroutines
are called immediately before the acraneb2 call and ensure that the 
largest arrays are moved the memory heap and that these arrays are 
properly NUMA-initialised, respectively. Correspondingly, two calls are 
made to the subroutines \verb|radia_dwarf_heap| and 
\verb|radia_dwarf_numainit| from \verb|<dwarf>/src/main.f90| in the 
subroutine structure that surrounds the radiation routine (see section 
\ref{sec:wrap}).

A detailed description of this dwarf prototype can be found in the report 
\cite{Poulsen_et_Berg_2017}. Here we will only add a summary of the 
steps taken in order to refactor the code:

\begin{enumerate}
  \item Establish a solid reference (test case and source code) that 
        reproduces the necessary results.
  \item Establish build and run environment to ease repetition and 
        reproducibility.
  \item Ensure proper threading, i.e.~a SPMD approach.
  \begin{itemize}
    \item This requires a transition to Fortran90 assumed shape and 
          trimming stack memory usage, and
    \item contiguous data
  \end{itemize}
  \item Strive towards a minimal implementation, including:
  \begin{itemize}
    \item Reducing the memory overhead.
    \item Reducing the stack pressure by reducing local 2D/3D variables 
          to 1D/2D vars or even scalars.
    \item The largest stack arrays should be moved to the heap with 
          proper NUMA initialisation of these heap arrays.
    \item Collapsing loops over the outermost index.
    \item Symbolic algebraic reduction using pen\&paper.
    \item Assuring no side effects in local functions (pure in Fortran).
    \item Declaring constants as constants (parameter in Fortran), not 
          as variables.
    \item Moving all branching out of the loops.
  \end{itemize}
  \item Continued refactoring by shuffling computations around to 
        maximize parallel exposure 
        (playing with data structures and loops).
  \begin{itemize}
    \item Identifying computational patterns, 
          e.g.~with reduction and prefix sums, and
    \item \ldots without (SIMD-suitable loops) dependencies.
    \item Re-organising heavy loops to constant trip-counts.
  \end{itemize}
\end{enumerate}

This list is taken from the presentation: ``Performance studies 
(ACRANEB2)'' by Per Berg and Jacob Weismann Poulsen given at the ESCAPE 
young scientist summer school in Copenhagen, August 2017. The full 
presentation is available from: http://www.hpc-escape.eu/media-hub/escape-events/ysss2017.

The module \verb|<dwarf>/src/dmi_omp.f90| has been added. From 
this the subroutine \verb|domp_get_domain| is called in order to 
balance the thread load based on local properties of ACRANEB2
\cite{Poulsen_et_Berg_2017}.

Of more basic structural changes a main thing is that the subroutine 
\verb|acraneb_transt| has been divided into three seperate subroutines: 
\verb|acraneb_transt1|, \\ 
\verb|acraneb_transt2| and 
\verb|acraneb_transt3|, which primarily cover the computations of 
descending optical depths (1), ascending optical depths (2) and the 
intermediate optical depths for the thermal radiative exchange between 
layers (3). Of these \verb|acraneb_transt3| has been found to be the 
most time-consuming subroutine and therefore particular focus has been 
given to this in the refactoring process. In this the loop structure 
has been reduced to a triangular loop structure in the refactoring 
process \cite{Poulsen_et_Berg_2017}. The \verb|delta_c_sca| and 
\verb|delta_t_sca| subroutines within \verb|acraneb_transt| have been 
rearraged in 6 differenct functions: \verb|zcdelta1|, \verb|zcdelta2|, \verb|zcdel0|, \verb|ztdelta1|, \verb|ztdelta2| and \verb|ztdel1|. This 
done to remove conditional statements at this deeply nested level of 
the code.

Other structural changes include that the subroutine 
\verb|ac_cloud_model2_t| now have replaced the \verb|ac_cloud_model2| 
subroutine. In the refactored version the number of output arrays is 
reduced, since these were only local arrays that could be calculated more 
efficiently at a later point in the code. The subroutines 
\verb|acraneb_coeftv0| and \verb|acraneb_coeftv1| replace the subroutine
\verb|acraneb_coeft|. With two subroutines replacing one subroutine 
conditional statements are again avoided.

\paragraph{Transt3 dwarfs for GPUs and Xeon Phi processors}

The dwarf version \inlsh{dwarf-transt3\_v0.1} is a stand-alone version 
of the \verb|transt3| subroutine described above (section 
\ref{sec:refactored}). Given that this subroutine takes more than 80\% 
of the total running time of ACRANEB2 \cite{Poulsen_et_Berg_2017}, 
it has been singled out for the optimised test on GPUs. This version 
is made for direct comparisons with the GPU-optimised dwarf versions 
\inlsh{dwarf-transt3\_gpu\_v0.2} and 
\inlsh{dwarf-transt3\_gpu\_nproma\_v0.1}. The ``nproma'' version of 
the GPU-optimised dwarf is hard-coded to run with \verb|NPROMA| set to 
a length of 32, since this was found to be optimal for the P100 GPU 
that has been tested \cite{Poulsen_et_Berg_2017}.

Binary input for these three dwarf versions are available in the file \\
\verb|escape-acraneb2-dwarf/test/escape-acraneb2-input/transt_in.bin|.

\subsubsection{Dwarf installation and testing}
\label{sec:installation}

The first step is to download and install the dwarf along 
with all its dependencies. With this purpose, it is possible 
to use the script provided under the ESCAPE software collaboration 
platform:\\
\url{https://git.ecmwf.int/projects/ESCAPE}.

Here you can find a repository called \inlsh{escape}.
You need to download it. There are two options to do this. One option is to use ssh. For this option you need to add an ssh key to your bitbucket account at \url{https://git.ecmwf.int/plugins/servlet/ssh/account/keys}. The link "SSH keys" on this website gives you instructions on how to generate the ssh key and add them to your account. Once this is done you should first create a 
folder named, for instance, ESCAPE, enter into it 
and subsequently download the repository by using the following the steps below:
\begin{lstlisting}[style=BashStyle]
mkdir ESCAPE
cd ESCAPE/
git clone ssh://git@git.ecmwf.int/escape/escape.git
\end{lstlisting}
The other option to download the repo is by using https instead of ssh. Instead of the git command above you then need to use 
\begin{lstlisting}[style=BashStyle]
git clone https://<username>@git.ecmwf.int/scm/escape/escape.git
\end{lstlisting}
where <username> needs to be replace by your bitbucket username.

Once the repository is downloaded into the \inlsh{ESCAPE} folder 
just created, you should find a new folder called \inlsh{escape}. 
The folder contains a sub-folder called \inlsh{bin} that has the 
python/bash script (called \inlsh{escape}) that needs to be 
run for downloading and installing the dwarf and its dependencies. 
To see the various options provided by the script you can type:
\begin{lstlisting}[style=BashStyle]
./escape/bin/escape -h
\end{lstlisting}
To download the dwarf you need to run 
the following command:
\begin{lstlisting}[style=BashStyle]
./escape/bin/escape checkout dwarf-P-radiation-ACRANEB2 \ 
--ssh
\end{lstlisting}
To use https you need to replace --ssh with --user <username>. The commands above automatically check out the \inlsh{develop}
version of the dwarf. If you want to download a specific branch 
of this dwarf, you can do so by typing:
\begin{lstlisting}[style=BashStyle]
./escape/bin/escape checkout dwarf-P-radiation-ACRANEB2 --user <username> \
--version <branch-name>
\end{lstlisting}
You should now have a folder called 
\inlsh{dwarf-P-radiation-ACRANEB2} with all 5 dwarf prototypes mentioned above inside of\\\inlsh{dwarf-P-radiation-ACRANEB2/sources/dwarf-P-radiation-ACRANEB2/src}. Each of these can then be configured and compiled with standard 
configure and make commands. Examples are given in the scripts 
\verb|<dwarf>/confmake.sh| for each of the dwarf versions. See also 
the README file: 
\verb|README-escape-acraneb2-dwarf.txt|.

\subsubsection{Running the Dwarf}
The ACRANEB2 prototypes are configured and made with the confmake.sh script.
On the KNL cluster cck at ECMWF you must first swap to the Intel environment: 

\begin{lstlisting}[style=BashStyle]
module swap PrgEnv-cray PrgEnv-intel
\end{lstlisting}

You can now use

\begin{lstlisting}[style=BashStyle]
cd <dwarf_directory>
./confmake.sh
\end{lstlisting}

and submit the dwarf to the queue with:

\begin{lstlisting}[style=BashStyle]
qsub submit.sh
\end{lstlisting}

To run the dwarf prototypes optimised for GPU processors on the GPU cluster lxg at ECMWF (K80 NVIDIA GPUs) you must first load and swap the 
modules:

\begin{lstlisting}[style=BashStyle]
module load cuda
module load openmpi
module swap gnu pgi
\end{lstlisting}

Then run: 

\begin{lstlisting}[style=BashStyle]
cd <dwarf_directory>
./confmake.sh
\end{lstlisting}

and submit the dwarf with:

\begin{lstlisting}[style=BashStyle]
sbatch submit_gpu.sh
\end{lstlisting}
\subsubsection{Conclusion remarks}

As this deliverable is being written, much work has already been done 
on the software adaptation (WP2) and benchmarking and diagnostics (WP3) 
for the ACRANEB2 dwarf. Based on this work, the following answers can 
be given to the questions posed in section \ref{sec:scope}:

\begin{enumerate}
  \item The potential of refactoring ACRANEB2 to run optimally 
        contemporary Xeon, Xeon Phi processors and GPUs is very large. 
  \item We find it very likely that this is the case for legacy physics 
        routines used in weather and climate models in general.
  \item For a complex physics subroutine, such as this, we found that 
        different refactoring was needed for the GPUs and the Xeon Phi 
        processors. 
  \item Rules can be made for how to adjust legacy physics code for 
        optimised running on current hardware architectures, however, the 
        work of specialists is required! Such an investment is 
        worthwhile given the significant improvement in 
        time-to-solution and reduction in energy consumption that can 
        be achieved.
\end{enumerate}